\documentclass[12pt,preprint]{aastex}

\slugcomment{To appear in ApJ}

\shorttitle{Lithium age for IC~2391}
\shortauthors{Barrado y Navascu\'es, Stauffer \& Jayawardhana }

\begin{document}

\title{Spectroscopy of Very Low Mass  Stars and Brown Dwarfs  in  
IC~2391:\\
Lithium depletion and H$\alpha$ emission}

\author{David Barrado y Navascu\'es}
\affil{Laboratorio de Astrof\'{\i}sica Espacial y F\'{\i}sica Fundamental,
INTA, P.O. Box 50727, E-2808 Madrid, SPAIN}
\email{barrado@laeff.esa.es}

\author{John R. Stauffer}
\affil{IPAC,  California Institute of Technology,
         Pasadena, CA 91125, USA }
\email{stauffer@ipac.caltech.edu}

\and

\author{Ray Jayawardhana}
\affil{Department of Astronomy, University of Michigan, 
830 Dennison Building, Ann Arbor, MI 48109, USA}
\email{rayjay@umich.edu}


%
%
%
\begin{abstract}
We have obtained intermediate-resolution optical spectroscopy of 44 
candidate very low mass members of the nearby young open cluster 
IC~2391. Of these, 26 spectra 
are totally new, 14 were already analyzed in a previous
paper and another four are in common. 
 These spectra, taken at the Cerro Tololo 4-meter and Magellan I 
and II telescopes, allow us to confirm 33 of them as likely cluster 
members, based on their spectral types, presence of Li, and H$\alpha$ 
emission. 
Among these new cluster members is CTIO-160 (M7), the first IC~2391 
candidate to satisfy all criteria for being a substellar member
of the cluster, including detection of the Li 6708A doublet.
With the enlarged membership, we are able to locate the lithium
depletion boundary of the cluster more reliably than in the
past.  Based on comparison to several theoretical models, we 
derive an age of 50$\pm$5 Myr for IC~2391.  We also estimate new
ages for the Alpha Per and Pleiades clusters; our ages are
85$\pm$10 Myr and 130$\pm$20 Myr, respectively.  We derive
an estimate of the initial mass function of IC~2391 that extends
to below the substellar limit, and compare it to those of other
well-studied young open clusters.   The index of the power law
mass function for IC~2391 is $\alpha$=0.96$\pm$0.12, valid in the range
0.5 to 0.072 M$_\odot$.
\end{abstract}




%
%
%
%
%
%
%
%
%
%
%
%
%
\section{Introduction}

Hundreds of open clusters\footnote{A comprehensive
database has been collected by J.-C. Mermilliod and can be
found at http://obswww.unige.ch/webda/}
 are known in the Galaxy. However, few among them 
are well characterized (size, distance and reddening)
 and fewer still are close enough to allow their stellar 
population to be investigated in detail. Only 
the Pleiades, the Hyades, Alpha Per and a few other
clusters  have been systematically studied. Our goal is to 
add other clusters to this list, and IC~2391 has been 
the focus of some of our efforts.

IC~2391 is a young cluster with an estimated age, based on 
Main Sequence isochrone fitting, of 35 Myr (Mermilliod 1981).
It is one of the nearest clusters, with a
Hipparcos distance modulus of  $(m-M)_o$=5.82$\pm$0.07
(Robichon  et al. 1999).
The interstellar reddening in its direction is very low,
E($B-V$)=0.04 or 0.06, as estimated by 
 Becker \& Fenkart (1974) and
Patten \& Simon (1996),  respectively. This last work also estimated
a distance of $(m-M)_o$=5.95$\pm$0.10, which is
the value we will use here. 

In Barrado y Navascu\'es,  Stauffer and Patten (1999, hereafter Paper I),
we presented some spectra of very low mass members of the cluster
and a preliminary age estimate, 53 Myr, based on the lithium depletion 
technique (see Stauffer et al. 1998; Basri, Marcy \& Graham 1996).
Subsequently, in  Barrado y Navascu\'es et al. (2001a, hereafter Paper II),
we conducted an extensive photometric survey in the optical
which yielded a substantial population of candidate members, both
 very low mass stars and brown dwarfs. The combination of this database 
with infrared photometry from 2MASS (Skrutskie et al. 1997, see also
 Cutri et al. 2003)  allowed us to extract from the initial sample 
those objects which might be interlopers, based on the analysis 
of different color-magnitude and color-color diagrams.
In the introduction to Paper II, we described  the results
achieved by previous studies, both those papers that looked for 
new cluster members, more massive than those presented in Paper II, and 
those papers that studied the properties of true cluster members
 (X--ray emission, rotation,  lithium abundance and so on). 
Since then, the only other works that
have been published dealing with properties
of members of this cluster are Randich et al. (2001) --spectra--
 and Allen et al. (2003) --the luminosity function and the age.

Here, we present  medium resolution optical spectra of a large 
number of candidate members discovered in Paper II and
consider the membership of those candidates by studying  their spectral types,
H$\alpha$ emission, sodium  and lithium content; derive again, in a 
more accurate way thanks to the larger number of objects, the location
of the lithium depletion boundary of the cluster and hence its age; and study
the Initial Mass Function (IMF).


%
%
%
%
%
%
%
%
%
%
%
%
%
\section{Observations}

%
%
%
\subsection{Sample.}

The spectra presented in this paper correspond to IC~2391 candidate
members discovered by two different groups.
The first (generally brighter) set was selected from 
Patten \& Simon (1996) and Patten \& Pavlovsky (1999). 
The second group was extracted from Paper II, where we presented a large sample of 
low mass stars and brown dwarf candidates in the range 12 $<$ $I_C$ $<$ 21,
discovered in a deep optical and infrared photometric search.
An initial subset of this survey was previously published in
Paper I, and it includes 
medium resolution spectroscopy, allowing the confirmation of the 
membership of most of them. These spectra were obtained
in January 1999 with the 
Ritchey-Chr\'etien spectrograph at the 
Cerro Tololo Interamerican Observatory (CTIO) 4 m telescope 
and have a spectral resolution equivalent to
 2.7 \AA{} (see Paper I for details).

In the present study, we have carried out spectroscopic observations
at medium resolution for a total of 44 candidate members.
Of these, 26 have been observed for the first time,
while  18 were already analyzed in Paper I.
 Four out of these 18 have been reobserved.
We  especially selected
those objects located around the lithium depletion boundary
in a color-magnitude diagram
 (generally very cool stars with masses slightly larger than 
the substellar limit at 0.072 M$_\odot$).
Figure 1 displays the $I$ magnitude versus the ($R-I$) color
in the  Cousins system. Proposed IC~2391 
  members are included as crosses. 
Open circles represent those candidates with spectra (in
some cases at quite low signal-to-noise) from Paper I,
whereas candidates with  new spectroscopic data  are illustrated
with solid circles in the diagram.
An  empirical ZAMS from Barrado y Navascu\'es et al. (2001b) 
is displayed as a solid line, as well as 
several 50 Myr isochrones (Baraffe et al. 1998; Siess et al. 2000;
D'Antona \& Mazzitelli 1997; long-dashed, dotted and short dashed, 
respectively). The location
of the lithium depletion boundary (LDB) estimated in Paper I is also included.
Due to the lack of space and since the main goal of Paper I
was to establish the location of the LDB of the cluster,
we did not publish all the spetra of that sample. We now present
them in Figure 2. 

Table 1 lists all the targets observed both in Paper I 
--except the three non-members analyzed there--
and in this paper, and includes accurate position (from the 
2MASS survey, Cutri et al. 2003), optical as well as 
near infrared photometry, and information regarding how the spectra were  
collected (telescope, instrument and date).

\subsection{Multifiber spectroscopy from CTIO.}

On 1999 March  10th and 13th, we collected multifiber medium resolution
spectroscopy with the Hydra~II bench spectrograph at the CTIO 4m 
telescope under a shared risk program.
 The  grating KPGL-D was used with the OG-570 filter.
The achieved resolution was 2.7 \AA{ } as measured in ThAr comparison
spectra observed with the same set-up, and covering the 
spectral range 6300-8500 \AA.

The total exposure time was 11.5 hours, but
 we divided this time into 12 individual observations of 1 hour each (except
the last one, which lasted only 30 minutes) during these two nights.
Each exposure  was processed individually with the 
IRAF\footnote{IRAF is distributed by National Optical Astronomy
Observatories, which is operated by the Association of Universities
for Research in Astronomy, Inc., under contract to the National
Science Foundation, USA} package ``hydra''. We used dome flats.
 After the extraction of each 
1-D spectrum and its calibration in wavelength, we combined all the 
data corresponding to the same target into a final spectrum using a median
 algorithm in order to remove the hits by cosmic rays. No correction
for telluric lines or instrumental profile was carried out.
The final spectra can be seen in Figure 3.

\subsection{Spectroscopy from Magellan I \& II.}

During two different observing runs mainly devoted to other 
goals, we were able to complement the previous data with additional 
spectra collected at the Magellan 6.5m twin telescopes, 
located at Las Campanas Observatory. The first campaign
took place on 2002 December 11-14th. 
We used Magellan I and the MIKE echelle spectrograph.  MIKE
provides spectral resolution that is actually too high for
our purposes.
For this reason and in order to improve the final signal-to-noise ratio,
we degraded the resolution by rebinning the original data during the
read-out to 2  and 8 pixels in the spatial and spectral  directions,
respectively, thus obtaining  spectra with a resolution
of 0.55 \AA{ } and complete spectral coverage of 4500- 7250 \AA.
Four IC~2391 candidate member were observed with MIKE.  We show
the H$\alpha$ order from these spectra in Figure 4a.
The second run took place on 2003 March 11th, and five additional
candidates were observed, in this case with Magellan II and the B\&C 
spectrograph, using the 1200 l/mm grating, yielding a 2.3 \AA{ }
resolution. These last spectra can be seen in Figure 4b (probable 
non-members are not included in the diagram).
In most cases, we took three individual exposures of 1200 seconds
each, fully reduced each exposure separately, and then added together
the three resultant 1D spectra at the end.
Additional details can be found in  Barrado y Navascu\'es et al. (2004).
\section{Analysis}

Table 2 lists our results, including the derived 
spectral type, the measured H$\alpha$, sodium and  lithium 
equivalent widths at 6563, 8200 and 6708 \AA, respectively,  and the estimated 
effective temperature and lithium abundance.
We used calibrations by Bessell (1979), Leggett (1992) and
Basri et al. (2000), and the ($R-I)_C$
color and the spectral types for  the $T_{eff}$ determination, and 
curves of growth by Zapatero Osorio et al. (2002) for
the lithium abundance.
In the subsequent section we analyze and discuss these 
results.

\subsection{Membership.}

%
\subsubsection{Spectral types.}

For most of our targets, 
spectral types were derived by comparison with 
spectral standards and/or other cluster members whose spectral 
types were known, 
in a similar manner to Kirkpatrick et al. (1999)
and Mart\'{\i}n et al. (1999), using several spectral indices
defined in the red side of the optical spectrum.
These standards, whose spectral types go from
K7 to M8,  were observed with the same spectral set-up.
In the case of the echelle spectra collected with the MIKE 
spectrograph, we estimated the spectral types using the 
TiO band-head which starts at 7053 \AA. The depth of 
this feature is very sensitive to the effective temperature and 
it is an excellent spectral type indicator. In this case, in
addition to our 
IC~2391 candidates, we observed standards  of almost every spectral
sub-class in the vicinity of M5, with half a sub-class steps.
A visual inspection, comparing IC~2391 members with the standard stars and
verifying the spectral classification by direct comparison among the IC~2391
objects themselves, was also carried out. We believe that the
uncertainty of the IC~2391 spectral types we have derived
is of order half a subclass.

Five of our fainter targets ($I_C$$\sim$20 mag), observed with HYDRAII,
 have spectra with a very low signal-to-noise and we have not attempted 
spectral classification. 


 The spectral type of several of our  brighter candidates  is in strong 
disagreement with membership in the cluster,  since they do not
correspond to the value expected from the optical and infrared 
colors. This is the case of two  stars --CTIO-002 and CTIO-067--
of  K spectral type, probable  background giants. These two stars, 
as well as the fainter objects, have no detectable  H$\alpha$ in 
emission or the line is seen in absorption (see next subsection).

\subsubsection{H$\alpha$ emission.}

The strength of the H$\alpha$ emission can be used as a criterion to
establish the membership of a cluster candidate. 
As stated before, two 
objects of K spectral type, warmer than the expected values 
from their colors, lack H$\alpha$ emission and, therefore, 
they can be classified as likely non-members. 
The same argument is valid for three out of the five faint objects 
located at the end of the cluster sequence ($I_C\sim$20),
 since we would have expected, at least, 
some emission. Note, however, that this criterion is a statistical
one, and membership cannot be completely ruled out for
 these three objects or confirmed for the other two.
In any case, the data suggest that there is a strong pollution rate 
for this range, about 60\%.

The comparison of the H$\alpha$ distribution
 between clusters of different ages 
is displayed in Figure 5. For clarity, we have only included 
the cluster associated to the multiple star $\sigma$ Orionis (Wolk 1996)
and the Alpha Per cluster. 
The data come from
B\'ejar et al. (1999); Barrado y Navascu\'es et al. (2001c, 2002, 2003);
Zapatero Osorio et al. (2002), 
Prosser (1992, 1994); Prosser \& Randich (1998); Stauffer et al. (1999).
We note that the $\sigma$ Orionis cluster,
with an age close to 5 Myr, has a significant number of stellar and
substellar members with H$\alpha$ far beyond the limits of the figure 
and might contain about 20 \% of classical TTauri stars and 
substellar analogs (Barrado y Navascu\'es et al. 2003; 
Barrado y Navascu\'es \& Mart\'{\i}n 2003; Jayawardhana et al. 2003), 
which are characterized, among other things, by strong, asymmetric
 and broad H$\alpha$ emission lines.
We find that the distribution of H$\alpha$ emission
in IC~2391 is very similar to that for the somewhat older Alpha 
Persei cluster.

Only one star belonging to IC~2391 stands  out in the figure, CTIO-059.
This object is shown with the average value corresponding to the
two observation we carried out in January 1999 at CTIO, but in fact it was 
observed in two consecutive nights, and showed
very different values (49.5 and 18.8 \AA, see Paper I). Therefore, 
it seems that we detected a flare in that cluster member at that time.

Figure 6 displays the H$\alpha$ equivalent width versus time.
There might be some variability on short time scales, within
the same night. This variability could be related with rotation
(see, for instance, possible modulation due to rotation in
CTIO-038 or CTIO-074).
Variability on a longer time scale might be present too, but the
dataset is too sparse and we cannot confirm this at the
present time. In any case, it would be very interesting to
compare this information with light curves derived for the
same objects.

\subsubsection{Sodium doublet at 8200 \AA.}

We have also measured the equivalent widths of the sodium doublet at
8200 \AA. The strength of this alkali feature is sensitive to
gravity (see, for instance, 
Mart\'{\i}n et al. 1996). Since IC~2391 members are 
much younger than field objects of similar spectral types, and 
should have larger radii,  it is possible to use this characteristic
 as a youth indicator. All our targets but one (CTIO-046) have 
W(Na{\sc i}) in agreement with a young age and, therefore, with membership.
The large equivalent width measured in CTIO-046 indicates that it is a
more evolved object, which confirms our other indications that this star 
is probably not a member of the cluster.

\subsubsection{Lithium at 6707 \AA.}

Figure 7 and 8 display the area around the Li{\sc i} 6708 \AA{ } for all our
spectra. The vertical dashed line indicate the location of this feature.
In some cases, the signal-to-noise is indeed good, and lithium is unambiguously
detected; see, for example, the cases  of CTIO-145 and 038 (January 1999),
 CTIO-081 (March 1999), CTIO-195, 192 and 026 (December 2002) and 
CTIO-160 (March 2003). Other cases, such as the two spectra of CTIO-077 
(January and March 1999), are less certain. In any case, the new data 
allow us a significant improvement in the
 determination of the location of the LDB (Section 3.2).

\subsubsection{Confirmed members at the substellar limit.}

So far, including the data published in Paper I, we have 
collected medium resolution spectroscopy for 47 candidate members
(out of the 206 identified in Paper II and a handful of brighter objects). 
The membership of another three (CTIO-040, CTIO-094 and VXR~27) was
rejected in Paper I.
Our targets were selected from those classified as probable and
possible members in Paper II based on their optical and
infrared photometry (132 objects in total).
In that study, we estimated the pollution rate for 
that subsample --candidates identified for the first time in Paper II-- as 25\%.
Out of the 47 spectroscopically observed candidates
(including some VXRP and PP objects), nine 
have been classified as non-members 
(the membership of  three out of this nine, namely CTIO-040, CTIO-094 and VXR~27, was
rejected in Paper I)   and another five 
as possible members (four from the CTIO sample). 
Of them, 38 objects (the CTIO-xxx) were discovered  and presented 
in Paper II and another six objects --brighter-- 
come from previous studies 
(from Patten \& Simon 1996 and Patten \& Pavlovsky 1999).
 Therefore, taking into account these 38 CTIO objects
 --including  4 possible members in the CTIO sample-, and CTIO-040 and CTIO-094
--the non-members discussed in Paper I, 
 the pollution rate is in the range 20-30 \% for the CTIO sample,
 depending on how the possible members are counted. 
%

%
%
%
\subsection{Lithium depletion boundary and cluster age.}

%
\subsubsection{Lithium equivalent width versus magnitudes and colors.}

The initial lithium depletion boundary for IC~2391
was  located  at $I_C$=16.2$\pm$0.2 and
$(R-I)_C$=1.91 in Paper I. Figures 7 and 8
 contain the area around the 
Li{\sc i}6708 \AA{ } doublet for the bona fide members.
The  initial estimate  of the LDB location is confirmed, but
 there are two stars --PP07 and CTIO-206-- that show lithium
in their spectra despite the fact they are brighter than the LDB.
Figures 9a and 9b display color-magnitude diagrams using 
optical and infrared data. We have used $I_C$ from Paper II
in the y-axis  for the first case and $Ks$ from 2MASS
in the second. In both  cases these two stars are clearly above 
the LDB. Actually, CTIO-206, the fainter of these two, 
 might be a binary composed of two very low mass
stars of almost the same mass, which would solve the puzzle.
However, the other star,  PP07, is well above the LDB, by 
about 2 magnitudes. In both cases, the Na{\sc i}(8200) equivalent 
width is in agreement with a young object.

A similar situation has been discovered in the 
Pleiades cluster by Oppenheimer et al. (1997), and their interpretation was
that the two supposed Pleiades members
(HHJ339=HCG332 and HHJ409=HCG509)
 are, in fact, young interlopers
in the line of sight.
Recently,  Deacon \& Hambly (2003) have derived  
membership probabilities for them. Although they are very low
(P=0.155 and P=0.284, respectively), these values
 are not conclusive, and membership
cannot be completely excluded.

The alternative would be that there is 
a mechanism which can prevent lithium depletion during the 
pre-main sequence phase in M dwarfs. Note that some warmer
Alpha Per and Pleiades members (K spectral type) may inhibit 
lithium depletion
using a mechanism related to rotation or magnetic field strength
(see, for instance, Soderblom et al. 1993;
 Garc\'{\i}a L\'opez, Rebolo \& Mart\'{\i}n 1994;
Randich et al. 1998;
  D'Antona et al. 1998).
 However, our medium resolution spectrum of PP07 does not seem
to be broader than the rest (although it would have been very
 difficult to detect anything with a projected rotational velocity of
 less than $\sim$100 km/s), and its activity in H$\alpha$ is average
compared with other cluster members of the same color or spectral type.

Another, more speculative, possibility is that these two objects are,
in fact,  bona-fide very low mass  members  of the cluster, which
have recently swallowed a companion (a brown dwarf). 
The sudden additional mass accretion could 
explain their location in these color-magnitudes diagrams
and the strong lithium feature in the spectrum. This mechanism has been
invoked to explain the tendency among planet-harboring stars to be metal rich
(Santos et al. 2001; Gonzalez et al. 2001). Possible evidence of planetary 
engulfment 
has been presented by Israelian et al. (2001; 2003), although it has been 
called into question by others (Reddy et al. 2002).

In any event, Figure 9 clearly shows the lithium chasm (i.e., the lack of 
lithium  for late K  and early M in the cluster).
Figure 10 helps to determine with greater degree of accuracy the location
of the chasm and, therefore, its cool border, the LDB.
Objects with lithium are displayed with solid symbols (circles for
our data, squares for data from Randich et al. 2001), whereas open symbols
represent objects without it (triangles for our upper limits).
The LDB keeps its location at $(R-I)^{LDB}_C$$\sim$1.9. 
Our first detection of lithium
is located at $I_C$=16.286,
 whereas the last star without it has a  magnitude of
 $I_C$=16.144 ($Ks$=13.394 and  13.587, respectively, if 2MASS data
are used instead). Since the adopted distance and reddening are
$(m-M)_0$=5.95$\pm$0.10 and E($B-V$)=0.06, equivalent to A$_I$=0.112
and  A$_K$=0.021  (Rieke \& Lebofsky 1985), these values  yield 
M($I_C$)$^{LDB}$ =  10.15 and 
M($Ks$)$^{LDB}$ =  7.52. 
The distance yielded by Hipparcos would locate the LDB 0.13 mag 
fainter than these values.

\subsubsection{Lithium depletion and a new lithium depletion boundary age estimate.}

The evolution of the lithium depletion boundary and the lithium 
chasm with age, from the empirical point of view,  is illustrated
in Figure 11, where we show a comparison with the three clusters where
this type of data are available (IC~2391, Alpha Per and the Pleiades).
Additional information can be found in 
Rebolo et al. (1996);
Stauffer et al. (1998, 1999);
Mart\'{\i}n  et al. (1998);
Basri \& Mart\'{\i}n (1999).
Note that a detection of the LDB has been attempted in a fourth cluster,
NGC2547, by Oliveira et al. (2003, see also Jeffries et al. 2003), 
but they were not able to detect it unambiguously.

We have already determined the absolute magnitudes in the $I_C$ and
$Ks$ filters where the LDB appears in IC~2391.
Using bolometric corrections
of BC$_I$=0.070 (Bessell 1991; Comer\'on et al. 2000)
and BC$_K$=2.756 (Tinney et al. 1993; Leggett et al.  1996), the
LDB is located at  M(bol,$Ks$)$^{LDB}$= 10.229 and 
 M(bol,$Ic$)$^{LDB}$= 10.251, or M(bol)$^{LDB}$= 10.24 mag.

We have analyzed in the same way all the data available in the 
literature for Alpha Per and the Pleiades clusters:

- In the case of Alpha Per, taking into account a distance
modulus of $(m-M)_0$=6.23 and E($B-V$)=0.096, and assuming that the
LDB is defined by AP310, AP322 and AP300 (AP325 might be a binary based on its
location in the color-magnitude diagram, CMD), we derive M($I_C$)$^{LDB}$=11.42
and M($Ks$)$^{LDB}$= 8.31. In the same way as for IC~2391, 
M(bol)$^{LDB}$= 11.31 mag.

- For the Pleiades cluster, the LDB is defined by CFHT-Pl-09, CFHT-Pl-10, 
Roque~16, and Teide~2 (CFHT-Pl-13). 
Assuming $(m-M)_0$=5.60 and E($B-V$)=0.04 (Pinsonneault et al. 1998),
we derive 
M($I_C$)$^{LDB}$= 12.18,
M($Ks$)$^{LDB}$= 8.94, and
M(bol)$^{LDB}$= 12.14 mag.

In all cases, an error of 0.15 mag has been estimated for the 
location of the LDB boundary, taking into account distances, reddening,
and the gap between Li detection and non-detection.

With these values and the predictions of  theoretical models, it is 
possible to estimate the age of each cluster. Figure 12 displays
the absolute magnitudes of an object whose lithium has been depleted
almost completely (one percent  of the original lithium abundance)
 versus the time. The models correspond 
to Baraffe (priv.comm) --M($I_C$) and  M($Ks$)-- and
D'Antona \& Mazzitelli (1997) --M(bol).
Similar diagrams can be created using models by Burrows et al. (1997).
The ages derived are 52, 51 and 46 Myr for IC~2391, 
79, 89 and 79 Myr for Alpha Per, and
122, 124 and 153 Myr for the Pleiades.
That is, we estimate the plausible ages for these three clusters
as 50$\pm$5, 85$\pm$10 and 130$\pm$25 Myr, respectively.
Note, however, that Jeffries \& Naylor (2001) have reevaluated 
the error budget for them, both experimental and systematic, and
estimated that the errors can be significantly larger.
As an example, a different  I-band bolometric correction
 (Monet et al. 1992, using a relationship between $(V-I)$ and $(R-I)$ colors)
 for the LBD  in the Pleiades yields an age of 140 Myr instead 153 Myr.
If we had used Baraffe's models instead of those 
from D'Antona \& Mazzitelli (1997), the ages derived from the M(bol)
would have been 135 or 125 Myr for these two different bolometric corrections.
Moreover, the use of Hipparcos distance would have added about 3 Myr to
 the age derived for IC~2391.

Burke, Pinsonneault \& Sills (2004) have also revaluated the 
the ages for these three clusters plus NGC2547, and examined the errors
in the analysis. Their values are  55$\pm$6, 101$\pm$12 and 148$\pm$19 Myr
for  IC~2391, Alpha Per and the Pleiades
(48, 87 and 126 Myr when introducing an ad hoc offset to the
I-band bolometric correction).
Within the error bars, all these results agree with each other.

As stated in Stauffer et al. (1998; 1999),  
Barrado y Navascu\'es, Stauffer \& Bouvier (1998), and
Paper I
--see a summary in Barrado y Navascu\'es, Stauffer \& Bouvier (1999)--  
these ages are about a 50\% older than the values 
obtained by fitting isochrones to the upper part of the 
Main Sequence. Recently, Allen et al. (2003) have
derived an age for IC~2391 based on the luminosity function (LF) of
the cluster. They obtained a value, 35 Myr,  lower than the LDB age
and  identical to the main-sequence turn-off age (Mermilliod 1981).
 By fitting the data published in Paper II with
models of the LF, they argue that LF has a peak at M(I)=14-15 mag,
 which should be produced by the deuterium burning, and that the age
cannot be $\sim$50 Myr. Their result supports a recent claim by 
Song, Bessell and Zuckerman (2002), who state that the LDB age might 
 overestimate the real age for   young clusters.
However, the analysis by Allen et al. (2003) was carried out 
prior to our new spectroscopic data,
 which indicate that the pollution rate 
in that range ($I_C\sim$20) is very large ($\sim$60 \%). The alleged LF peak
is not obvious at all now. Additionally, Dobbie et al. (2002) have pointed out
that a drop in the LF exists around M7-M8 (Teff$\sim$2500 K), which
might be due to dust formation in the atmospheres of these objects. For 
IC~2391, this happens at about $I_C\sim$19.5 mag.

For these three clusters, the lithium depletion boundaries are
located at  the  spectral types of M5, M6.5 and M6.5
or, in effective temperatures, 3050, 2800 and 2650 K, for IC~2391,
 Alpha Per and 
the Pleiades, respectively. When expressed in  mass, 
using models from Baraffe et al. (1998), 
they take place at 0.12,  0.085 and 0.075  M$_\odot$, repectively.
All the LDB data are summarized in Table 3.

\subsubsection{The first confirmed brown dwarf in the cluster.}

So far, although a large number of candidate brown dwarfs
were presented in Paper II, none of these candidates were 
established via spectroscopy.
This confirmation implies that:
(a) it is a member of the cluster based on all available criteria;
(b) it has a bolometric magnitude that would place it below the
substellar mass limit if it were a member of the cluster; and
(c) there is a detection of the lithium 6708A doublet.

Using our new age estimate and 
models by  Baraffe et al. (2002), the interstellar reddening and the  
cluster distance, the substellar limit is located at
 $I_C$=17.06. Several of the targets in this sample
are fainter than this value, and their spectroscopic
properties agree with membership. However, their spectra are
not good enough to have a lithium detection beyond a doubt, which 
would confirm the substellar nature. Only in one case, CTIO-160 
(whose spectral type is  M7), is 
lithium clearly identified and its nature firmly established, 
making this object the first brown dwarf unambiguously identified 
in the IC~2391 cluster. Note, however, that errorbars in the object 
photometry and the uncertainties in the models are large enough
to change the classification of this object.

\subsection{Mass Function.}

We have derived a mass function for IC~2391 using 
non-dusty models  from Baraffe et al. (1998) and the $Ic$ magnitudes. Figure 13 depicts 
our results. We have assumed  different ages, ranging from 
25 to 50 Myr (these values are close to the turn-off and the LDB ages,
respectively). In any case, when expressed as a power law, 
the index ($\alpha$=0.96$\pm$0.12) does not depend  on the age in this interval
(i.e., the derived value is very similar when using these three ages).
The MF  is valid between a mass of 0.5 M$_\odot$ and the substellar limit.
Below 0.072 M$_\odot$, there is a sudden drop, which might be partially 
explained by the lack of survey completeness beyond $I_C$$\sim$18.5 for
 cluster members (0.050 M$_\odot$ for 50 Myr isochrone from Baraffe et al. 1998).
 However, we have detected a significant number of
 candidate members with magnitudes around $I_C$$\sim$20. Our
medium resolution spectroscopy indicates that, despite the strong
pollution in this range,  about  60 \%, some
seem to be members of the cluster. Therefore, the gap at 
$I_C$$\sim$19, mass$\sim$0.05  M$_\odot$ might be real.
Dobbie et al. (2002) and Jameson et al. (2003) have pointed out that several young clusters
show a lack of substellar members of M7-M8 spectral type, more or less 
in the same location as in the case of IC~2391 (see the case
of Alpha Per in the same diagram). They explain this
fact as an effect of dust formation  at this spectral type
and its effect in the luminosity function  as a new source of opacity, 
which would decrease the overall luminosity for the cooler objects.
In any event, the number of objects discovered so far in the IC~2391
cluster at the low end of the cluster sequence is too low 
to confirm this possibility.

Figure 13 also contains a comparison with several Mass Functions 
corresponding to  young  clusters of different ages, such as 
Alpha Per, the Pleiades, and M35 all of them derived in the same manner
(see Bouvier et al. 1998; Barrado y Navascu\'es et al. 2001b; 
 Barrado y Navascu\'es et al. 2002;
 Barrado y Navascu\'es \& Stauffer 2003; 
 Barrado y Navascu\'es \ 2003).
 The index of the Mass Function
power law is very similar in all cases, except in the case of 
the low mass stellar members of M35, a very rich cluster, where 
some mass segregation might have taken place due to its older age
(175 Myr in the turn-off age scale, Barrado y Navascu\'es et al. 2001d).

\section{Conclusions and Summary}

By collecting  medium resolution spectroscopy for a significant 
fraction of IC~2391 candidate members discovered in Paper II,
 we have established the membership
for most of them via their spectral types and H$\alpha$ emission
properties, including dependence with spectral type and 
variability in a short time scale.

In addition, we have studied the presence of the lithium 
doublet at 6708 \AA, located the lithium depletion boundary in 
the color-magnitude diagram and, 
with the help of theoretical models, derived an age estimate, 50$\pm$5 Myr.
The same study was carried out in other two clusters, namely Alpha Per
and the Pleiades. Our new age estimate is 85$\pm$10 and 130$\pm$20 Myr.
We have also derived an Initial Mass Function for the low mass end
of the IC~2391 cluster, fitting a power law with an index of 
$\alpha$=0.96$\pm$0.12.


%
%
%
%
%
%
%
%
%
%
%
%
%
\acknowledgements
We thank the staff of the CTIO 4m and Magellan telescopes
 for outstanding support. The anonymous referee has indeed contibuted
to the improvement of the paper.  DByN is indebted to the Spanish
``Programa Ram\'on y Cajal'', AYA2001-1124-CO2 \& AYA2003-05355 programs. 
R.J. acknowledges support from NSF grant AST-0205130. 
This publication makes use of data products from the Two Micron 
All Sky Survey. 

%
%
%
%
%
%
%
%
%
%
%
%
%


\newpage

\setcounter{figure}{0}
    \begin{figure*}
    \centering
    \includegraphics[width=16.2cm]{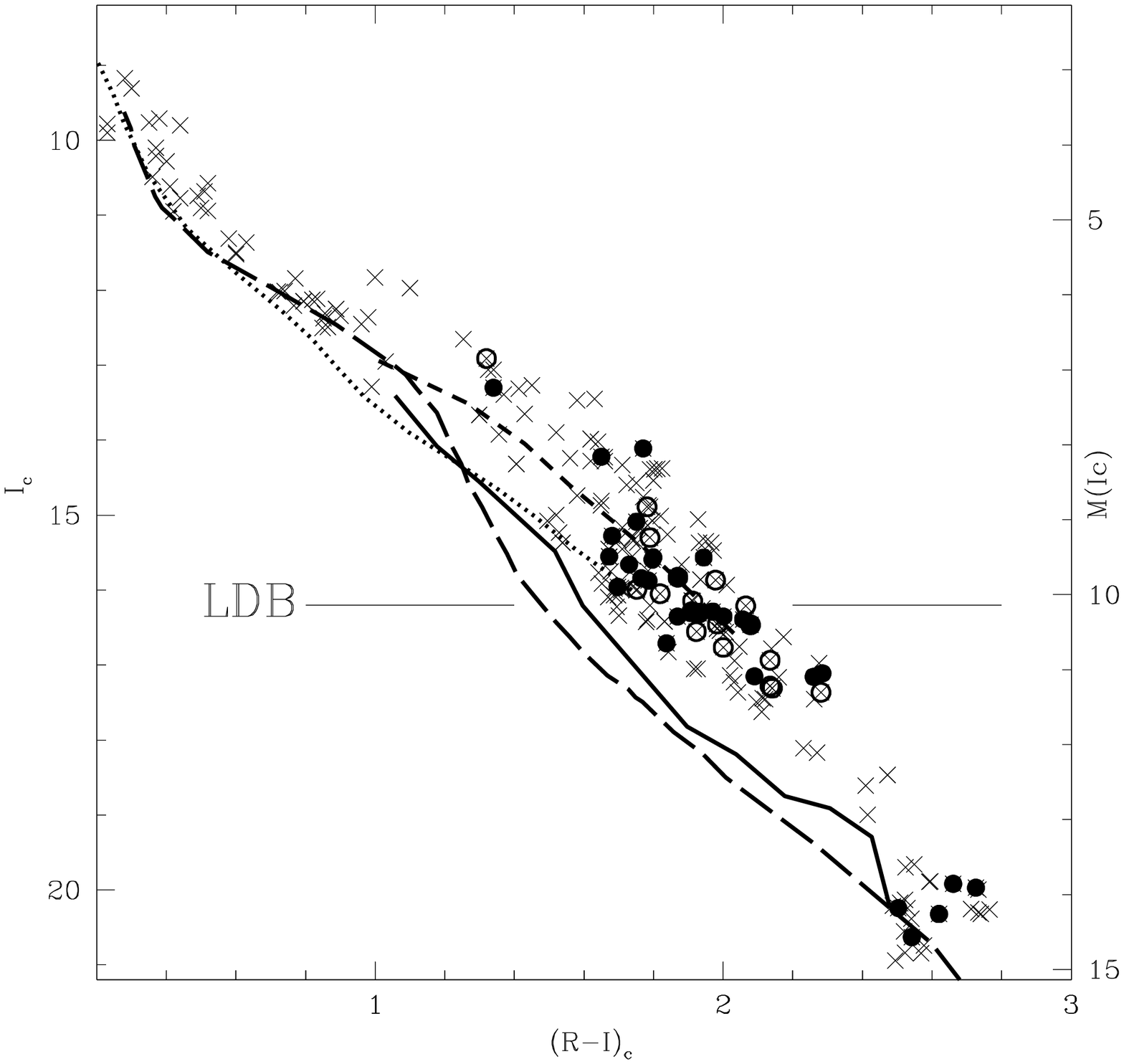}
 \caption{Color-magnitude diagram for IC~2391 candidate members.
Crosses represent all the available photometric data from 
Simon \& Patten (1996), Patten \& Pavlovsky (1999) 
and Barrado y Navascu\'es et al. (2001a).
Open circles correspond to spectroscopic data from 
Barrado y Navascu\'es et al. (1999), whereas solid circles
were observed with HydraII or the Magellan I and II telescopes.
We plot different 50 Myr isochrones (short dashed for
 D'Antona \& Mazzitelli 1997, long dashed for Baraffe et al. 1998 and
dotted for Siess et al. 2000). The solid line represents
 an empirical ZAMS.}
 \end{figure*}

\setcounter{figure}{1}
    \begin{figure*}
    \centering
   \includegraphics[width=8cm]{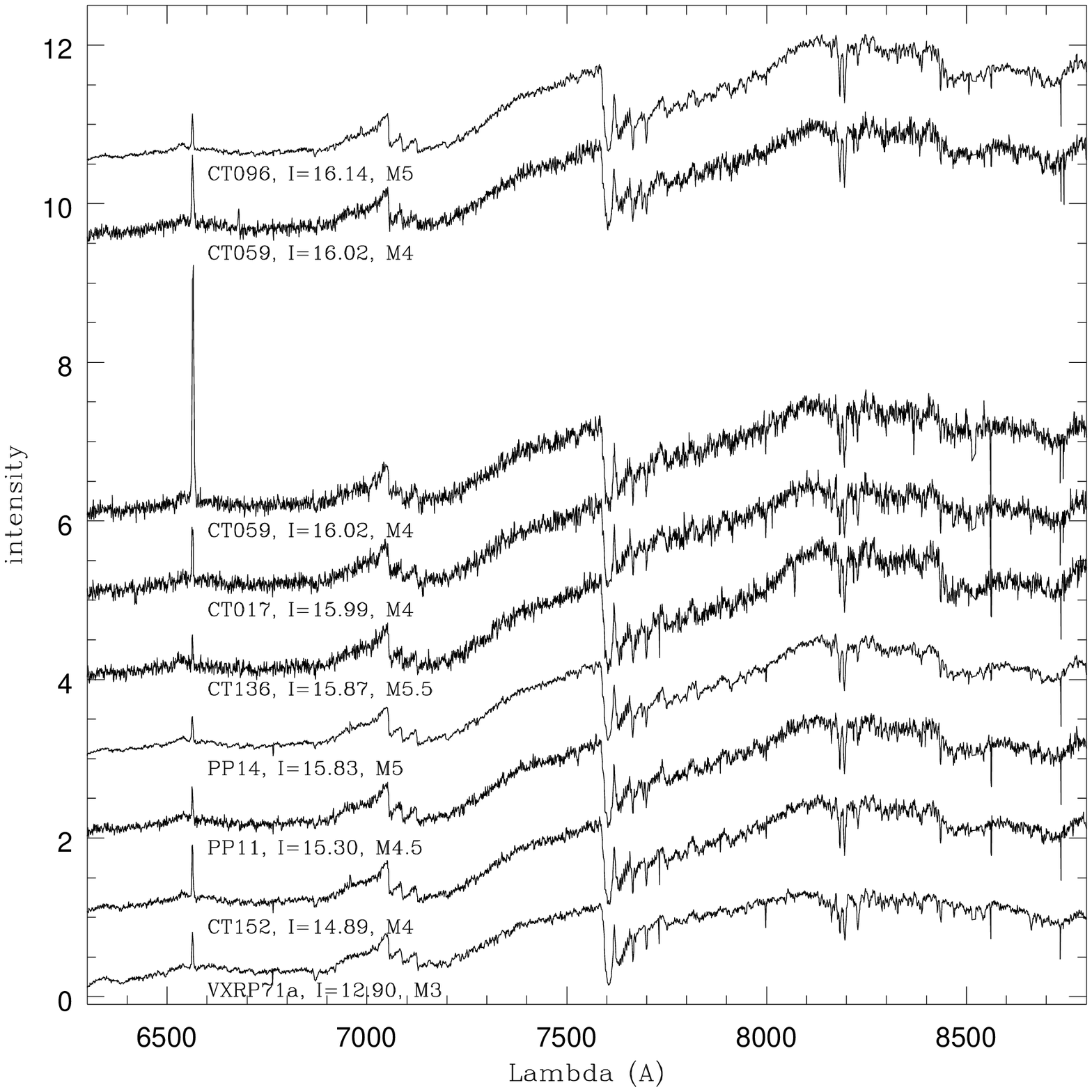}
   \includegraphics[width=8cm]{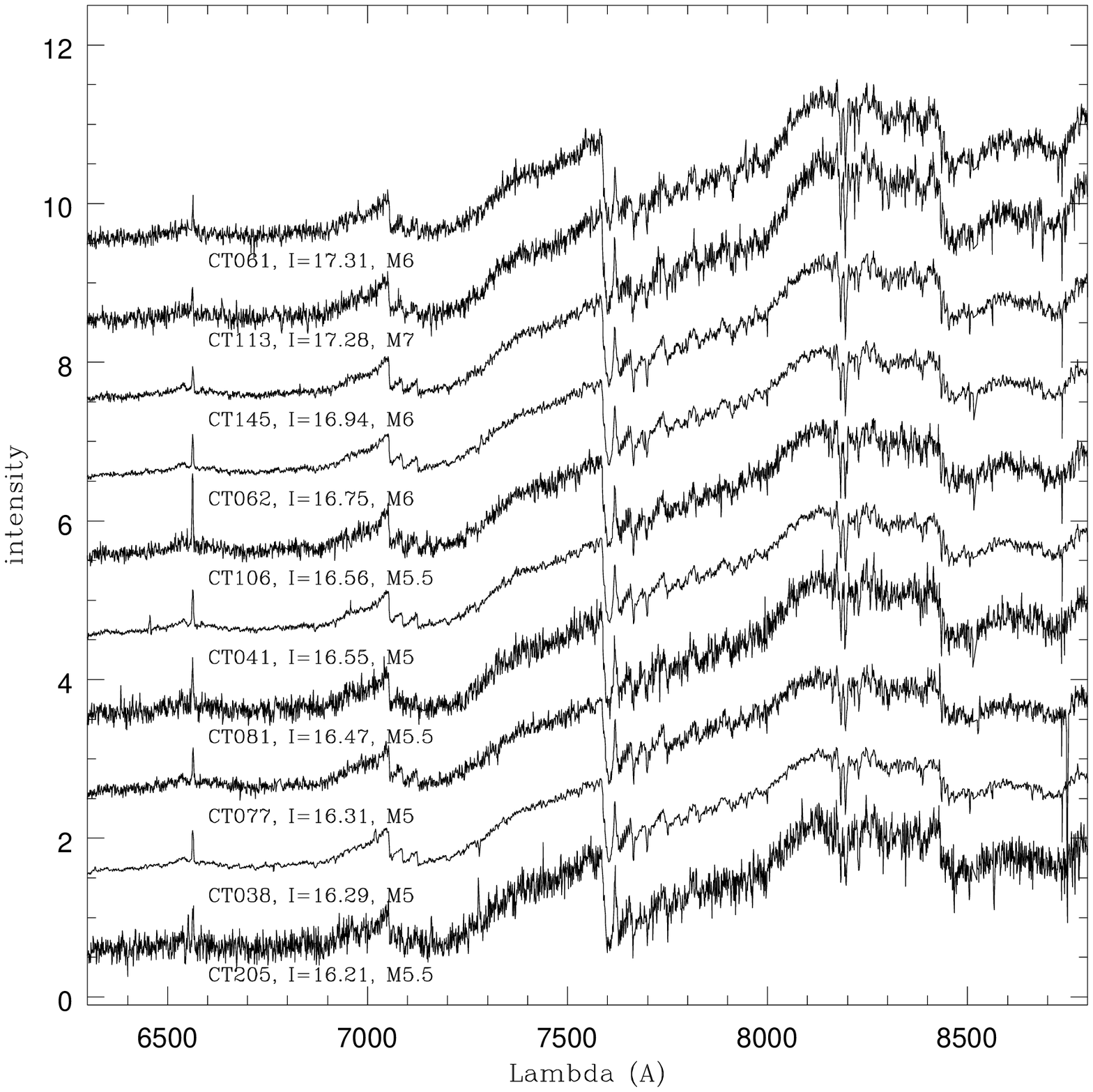}
 \caption{Spectra corresponding to our January 1999 run at CTIO, with
the Ritchey-Chr\'etien spectrograph. Note the change in the emission
of CTIO-059, probably due to a flare.}
 \end{figure*}

\setcounter{figure}{2}
    \begin{figure*}
    \centering
   \includegraphics[width=8cm]{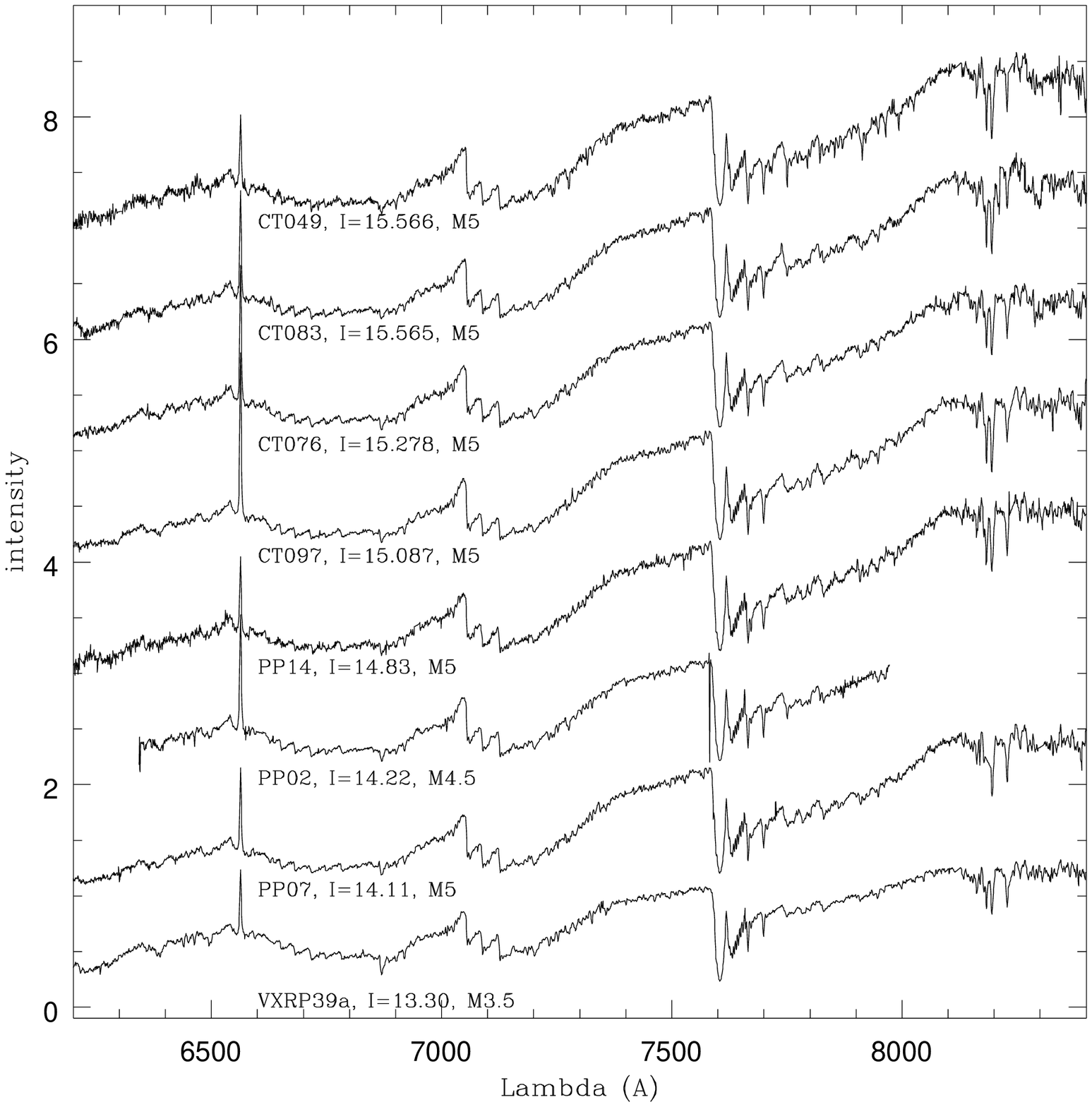}
   \includegraphics[width=8cm]{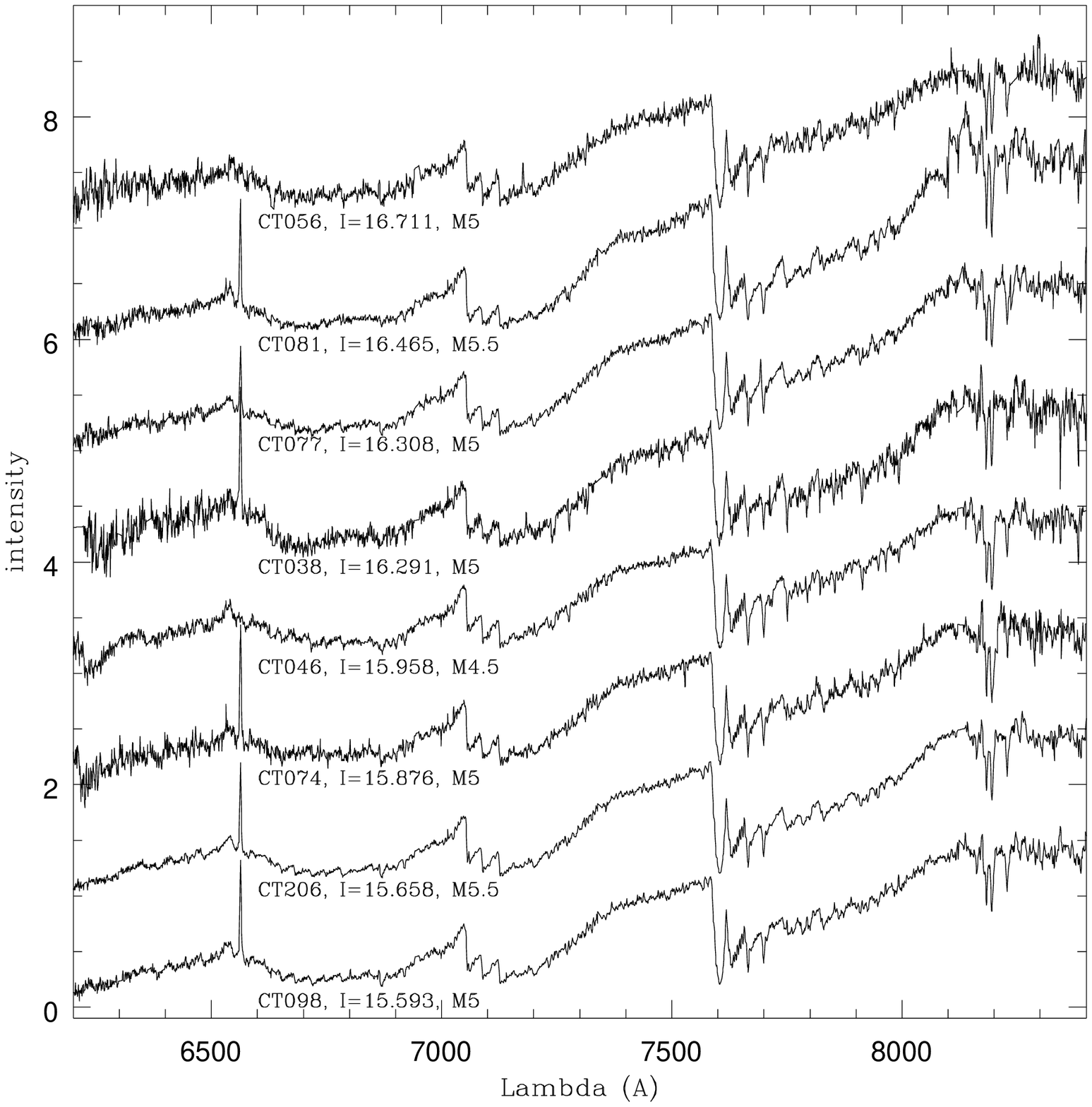}
 \caption{Spectra collected with CTIO$+$HydraII during our March 1999 run.
}
 \end{figure*}

\setcounter{figure}{3}
    \begin{figure*}
    \centering
   \includegraphics[width=8cm]{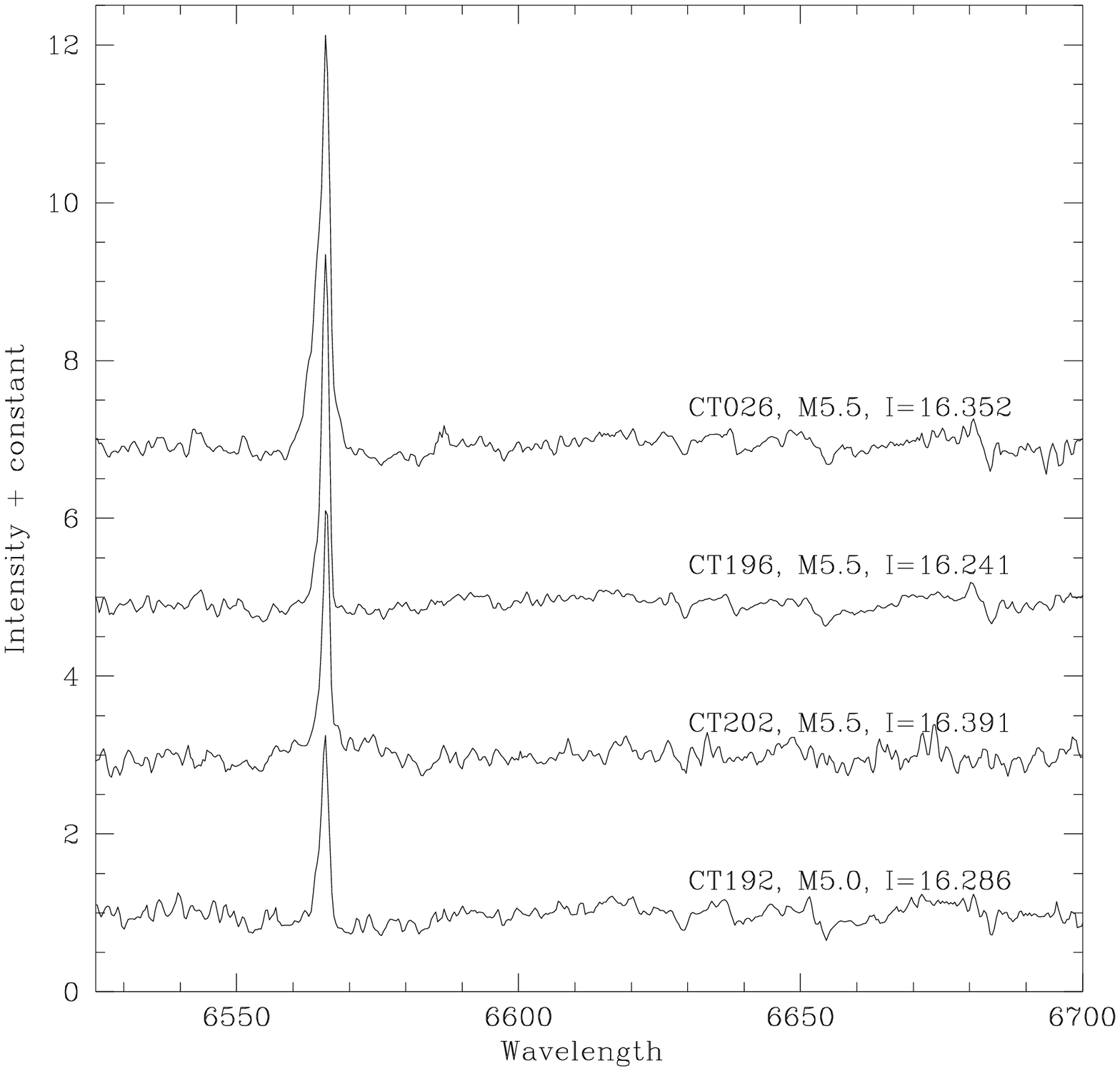}
   \includegraphics[width=8cm]{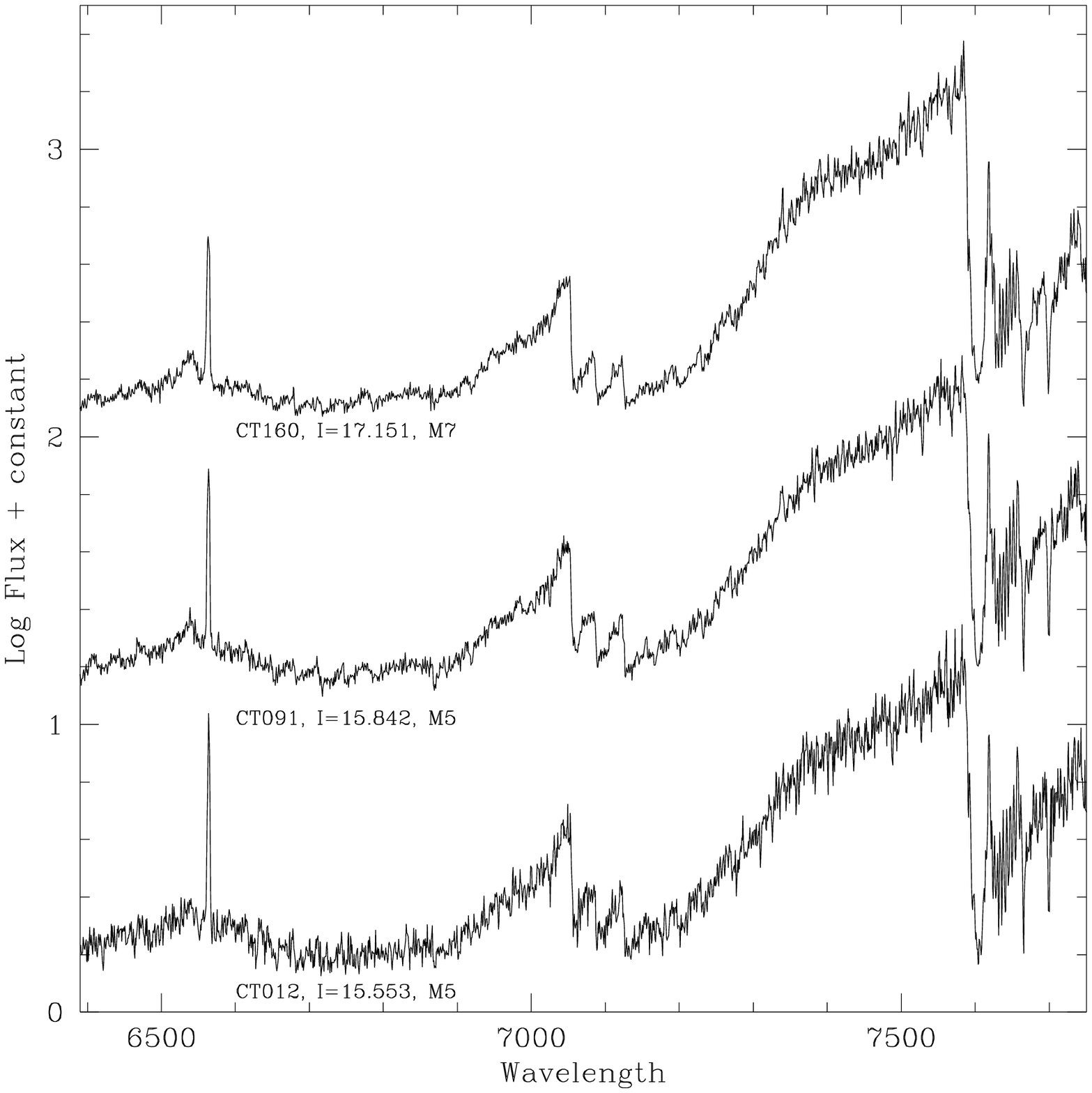}
 \caption{
{\bf a}
December  2002 spectra taken with the Magellan I
telescope and the MIKE echelle spectrograph. We only show the order
corresponding to the H$\alpha$ line. Note that 
the continuum has been normalized.
{\bf b}
March 2003 spectra (only members) taken with the Magellan II
telescope and the B\&C spectrograph.
}
 \end{figure*}

\setcounter{figure}{4}
    \begin{figure*}
    \centering
    \includegraphics[width=16.2cm]{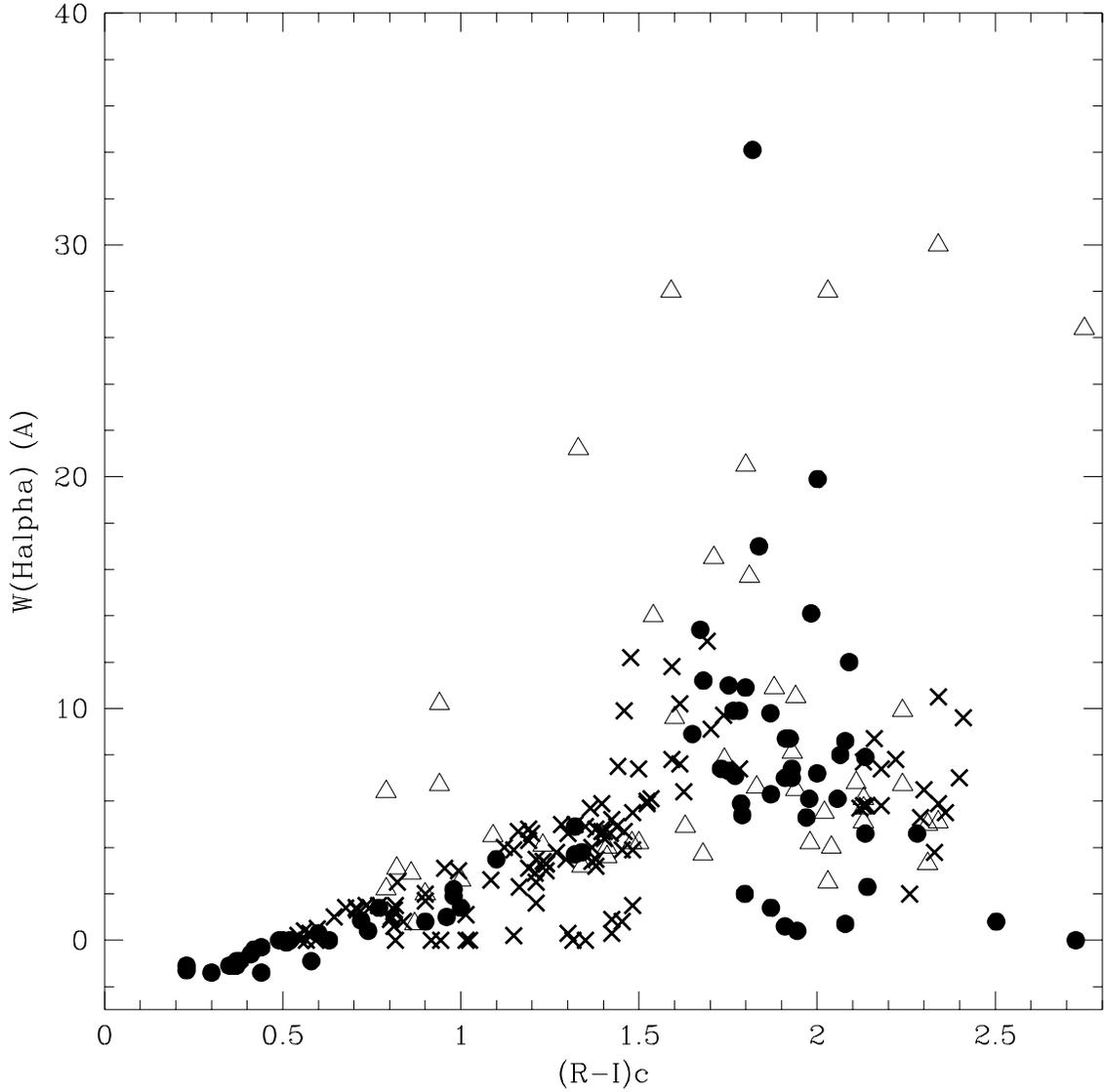}
 \caption{Comparison between the H$\alpha$ equivalent widths versus the 
$(R-I)c$ color index for several clusters. Triangles, circles, and crosses
represent data from Sigma Orionis, IC~2391 and Alpha Persei clusters, 
respectively. Note that a significant fraction of Sigma Orionis members 
(eight in this color interval) have H$\alpha$ larger than 40 \AA.
This is likely an affect of accretion by a disk (Barrado y Navascu\'es
et al. 2003; Barrado y Navascu\'es \& Mart\'{\i}n 2003). }
 \end{figure*}

\setcounter{figure}{5}
    \begin{figure*}
    \centering
   \includegraphics[width=10cm]{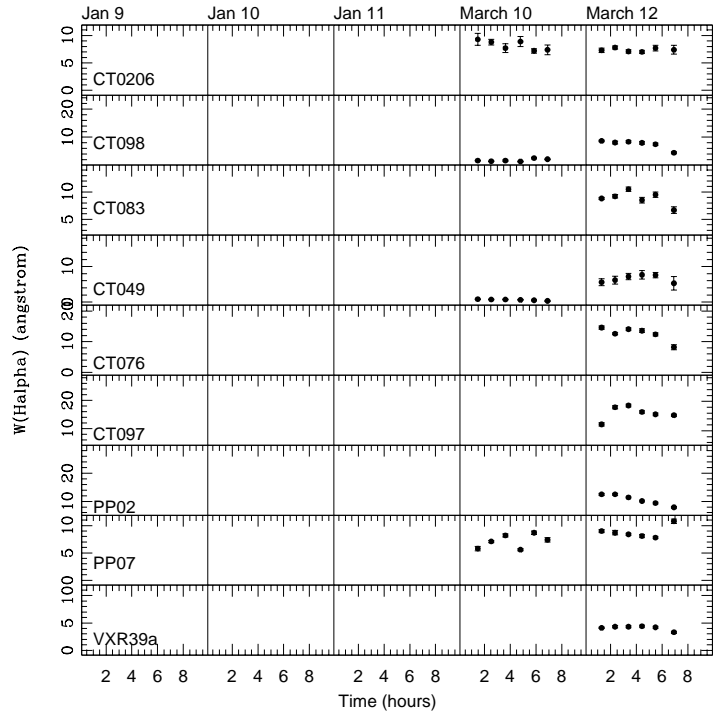}
   \includegraphics[width=10cm]{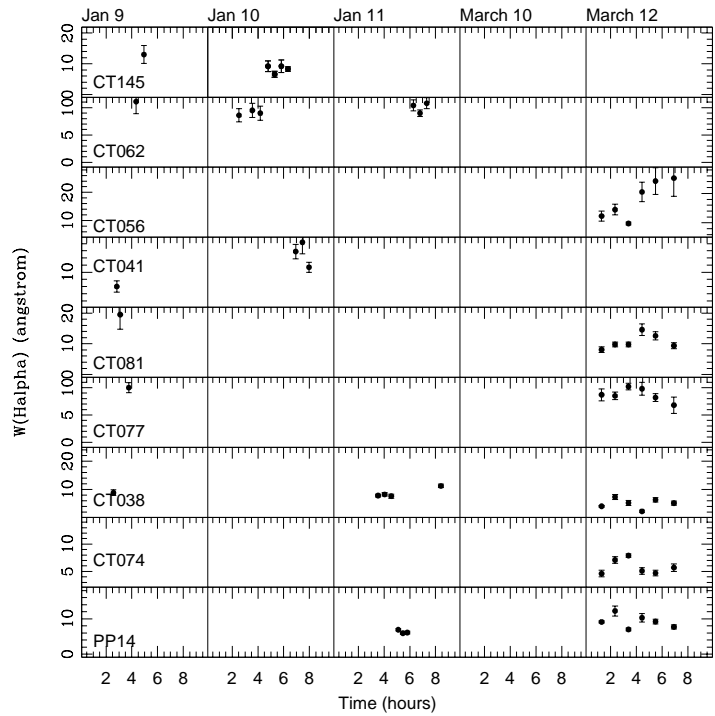}
 \caption{H$\alpha$ equivalent width versus time. 
The data correspond to the January and March 1999 campaigns.
Note the date on top of each panel.}
 \end{figure*}

\setcounter{figure}{6}
    \begin{figure*}
    \centering
   \includegraphics[width=8cm]{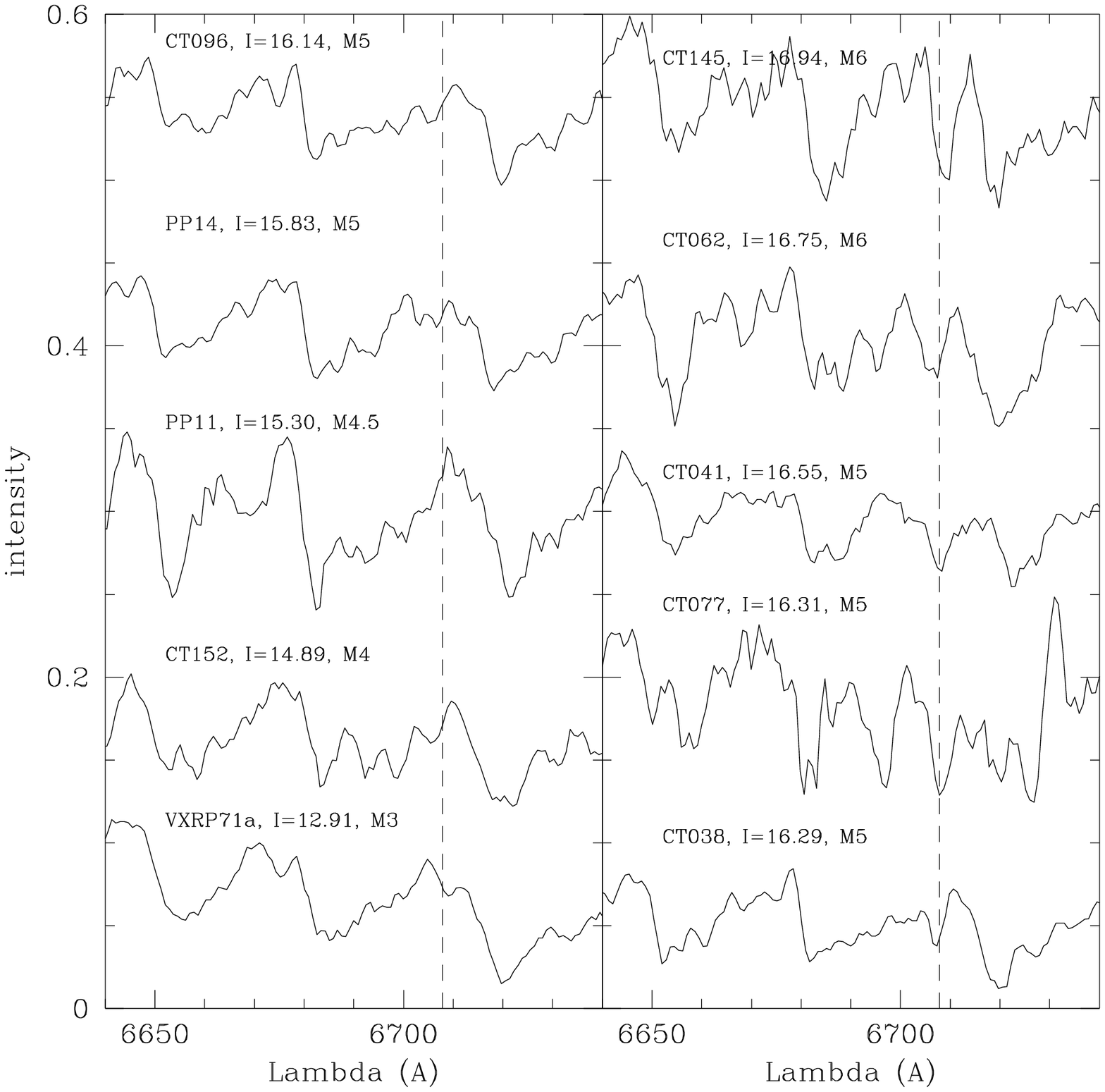}
   \includegraphics[width=8cm]{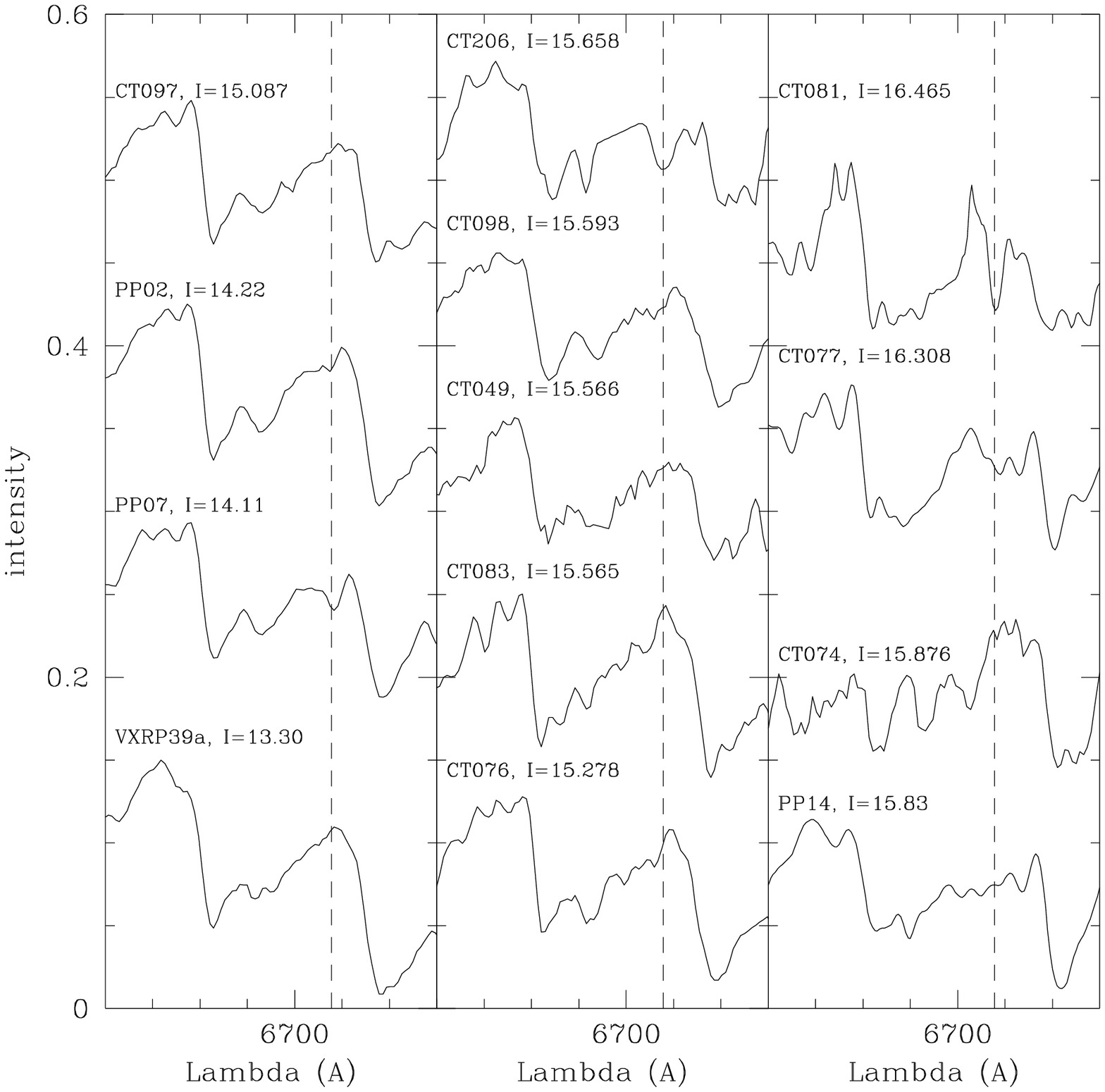}
 \caption{Detail  around the Li6708 \AA{} feature
for the January 1999 and March 1999 dataset (panels a and b).}
 \end{figure*}

\setcounter{figure}{7}
    \begin{figure*}
    \centering
   \includegraphics[width=8cm]{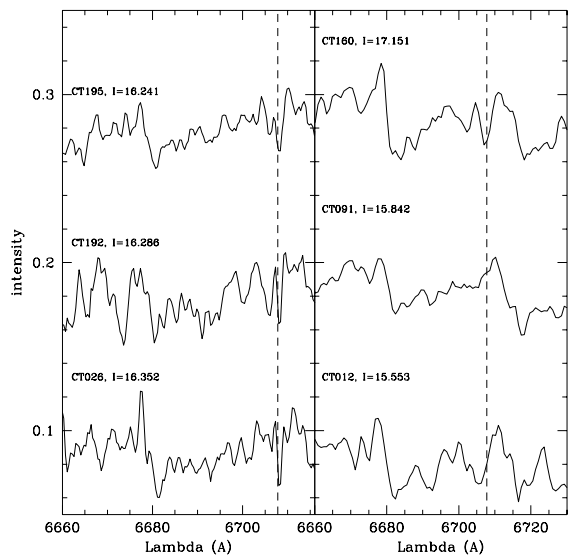}
 \caption{Detail of the spectra around the Li6708 \AA{} feature
for the December 2002  and March 2003 spectra 
(panel a and b, respectively, with resolutions  0f 0.55 and 2.3 \AA).}
 \end{figure*}

\setcounter{figure}{8}
    \begin{figure*}
    \centering
    \includegraphics[width=10.2cm]{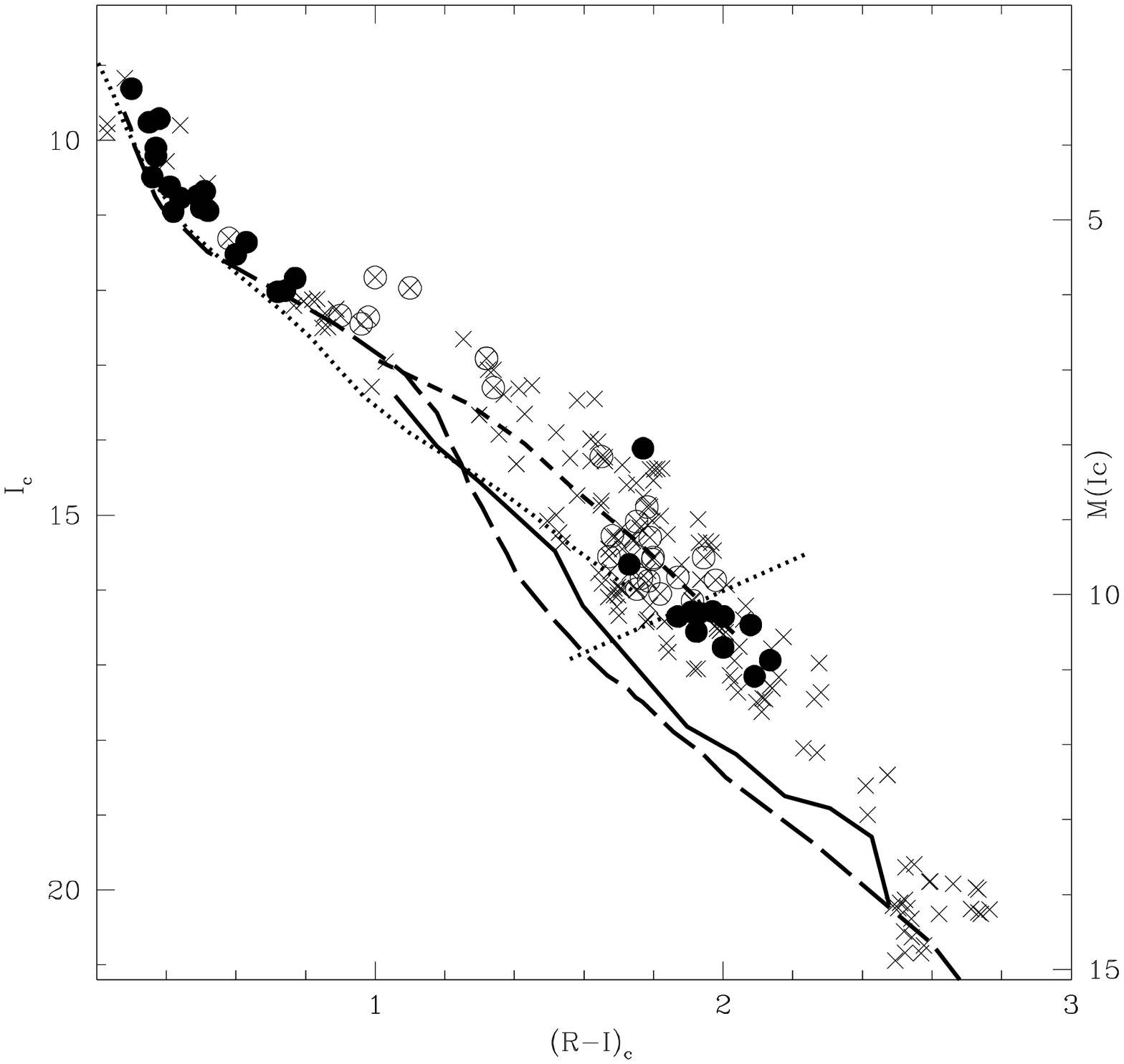}
    \includegraphics[width=10.2cm]{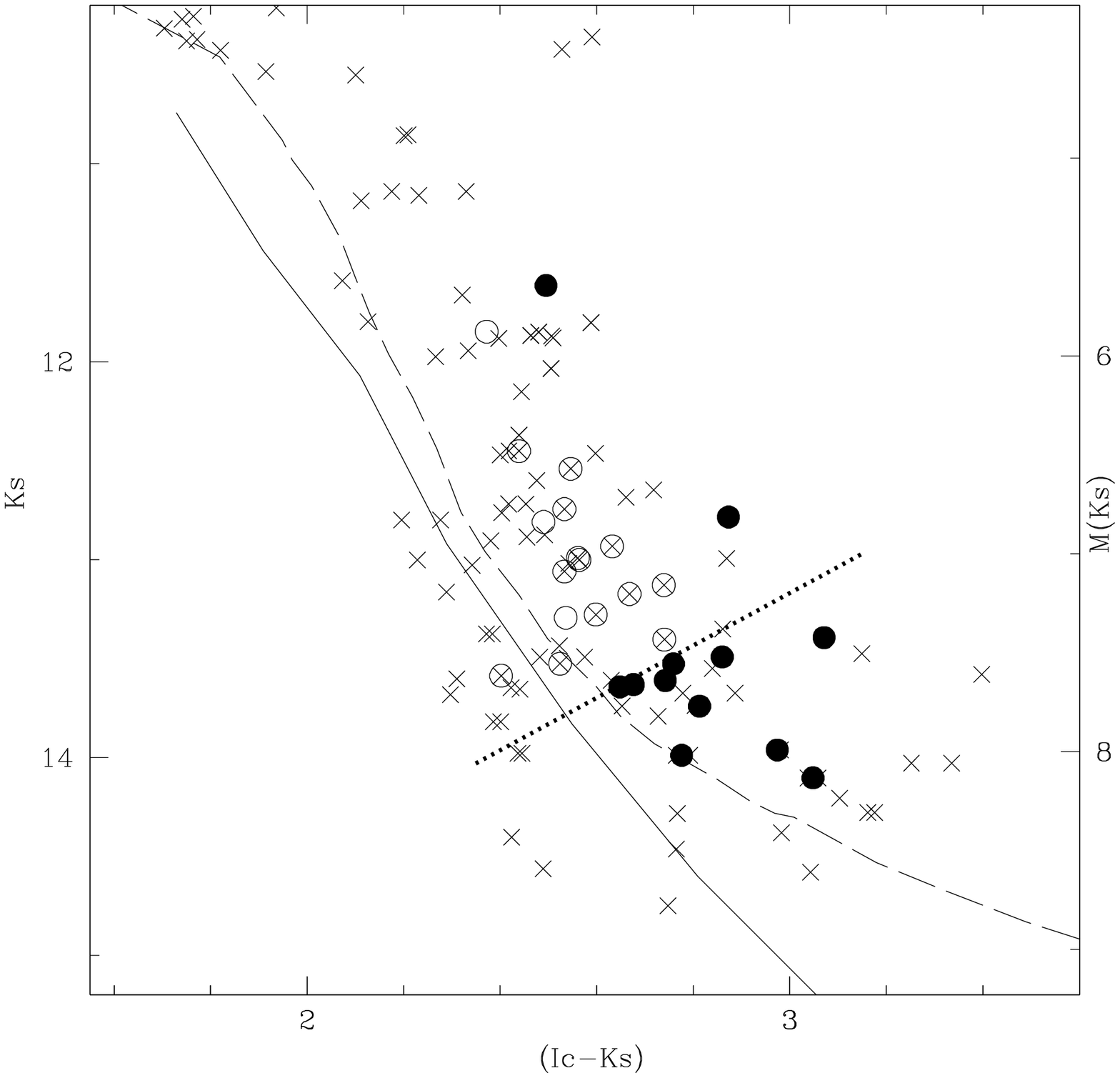}
 \caption{Color-magnitude diagrams and the LDB
 for IC~2391 candidate members.
Crosses represent all candidate members from 
Simon \& Patten (1998), Patten \& Pavlovsky (1999) 
and Barrado y Navascu\'es et al. (2001a).
IC~2391 members with lithium detection are
shows as solid circles, whereas those lacking the Li6708 \AA { }
feature are displayed as open symbols.
{\bf a}
We plot different 50 Myr isochrones --short dashed for
 D'Antona \& Mazzitelli (1997), long dashed for Baraffe et al. (1998) and
dotted for Siess et al. 2000). The dolid line represents
 an empirical ZAMS.
{\bf b}
A 50 Myr isochrones is also included (Baraffe et al. 1998, long-dashed line),
 as well as an empirical ZAMS (solid line) and the location of the
lithium depletion boundary (dotted line).}
 \end{figure*}

\setcounter{figure}{9}
    \begin{figure*}
    \centering
    \includegraphics[width=16.2cm]{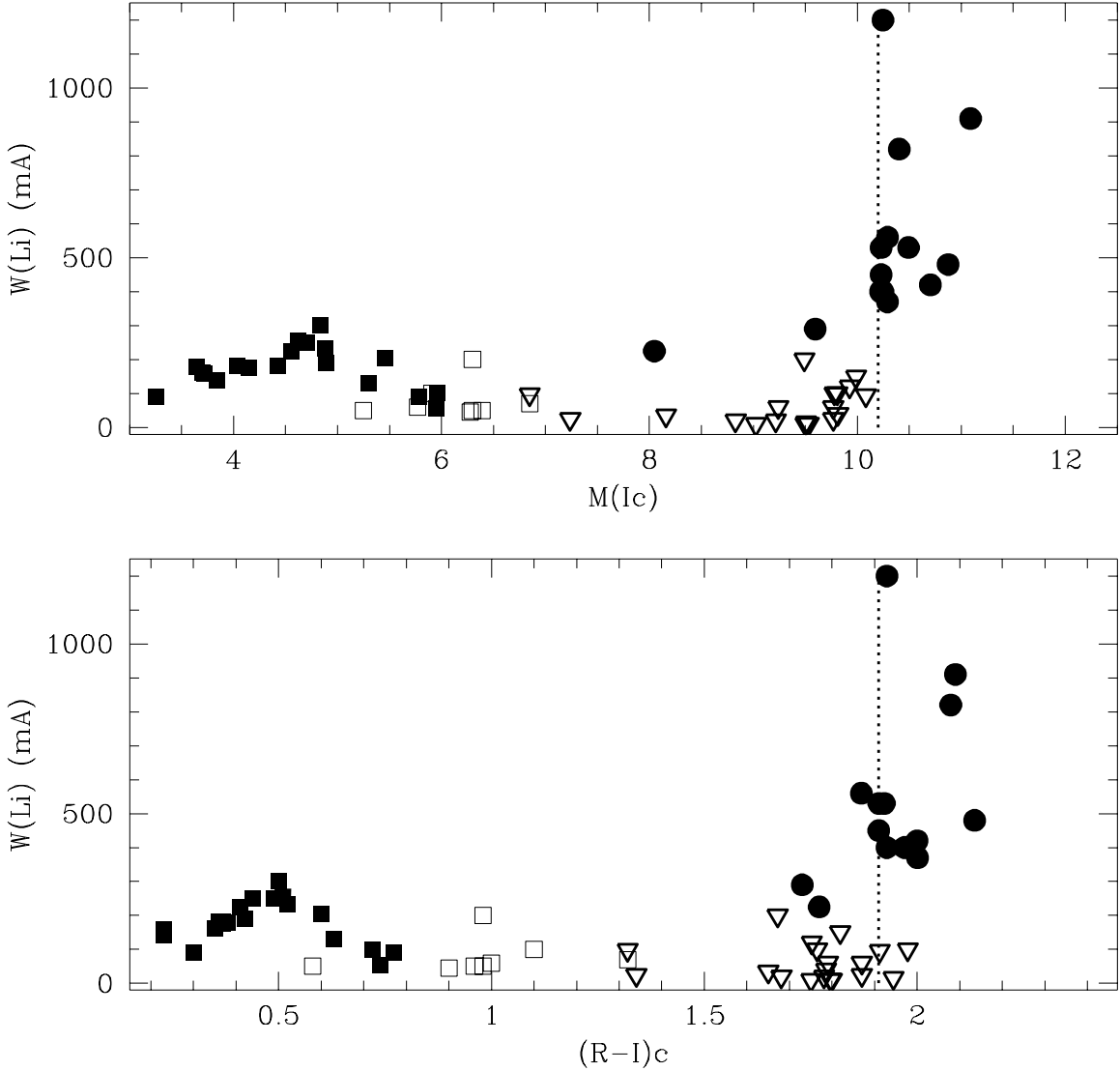}
 \caption{Lithium equivalent width versus the absolute $Ic$ 
magnitude ({\bf a}) and $(R-I)c$ color index  ({\bf b}).
Circles and triangles represent data from this work and 
Barrado y Navascu\'es et al. (1999), whereas squares correspond to data from
the literature. Actual data and upper limits  are displayed as solid  
and  open symbols, respectively.
The vertical dotted line locate the lithium depletion boundary
for the cluster.}
 \end{figure*}

\setcounter{figure}{10}
    \begin{figure*}
    \centering
    \includegraphics[width=16.2cm]{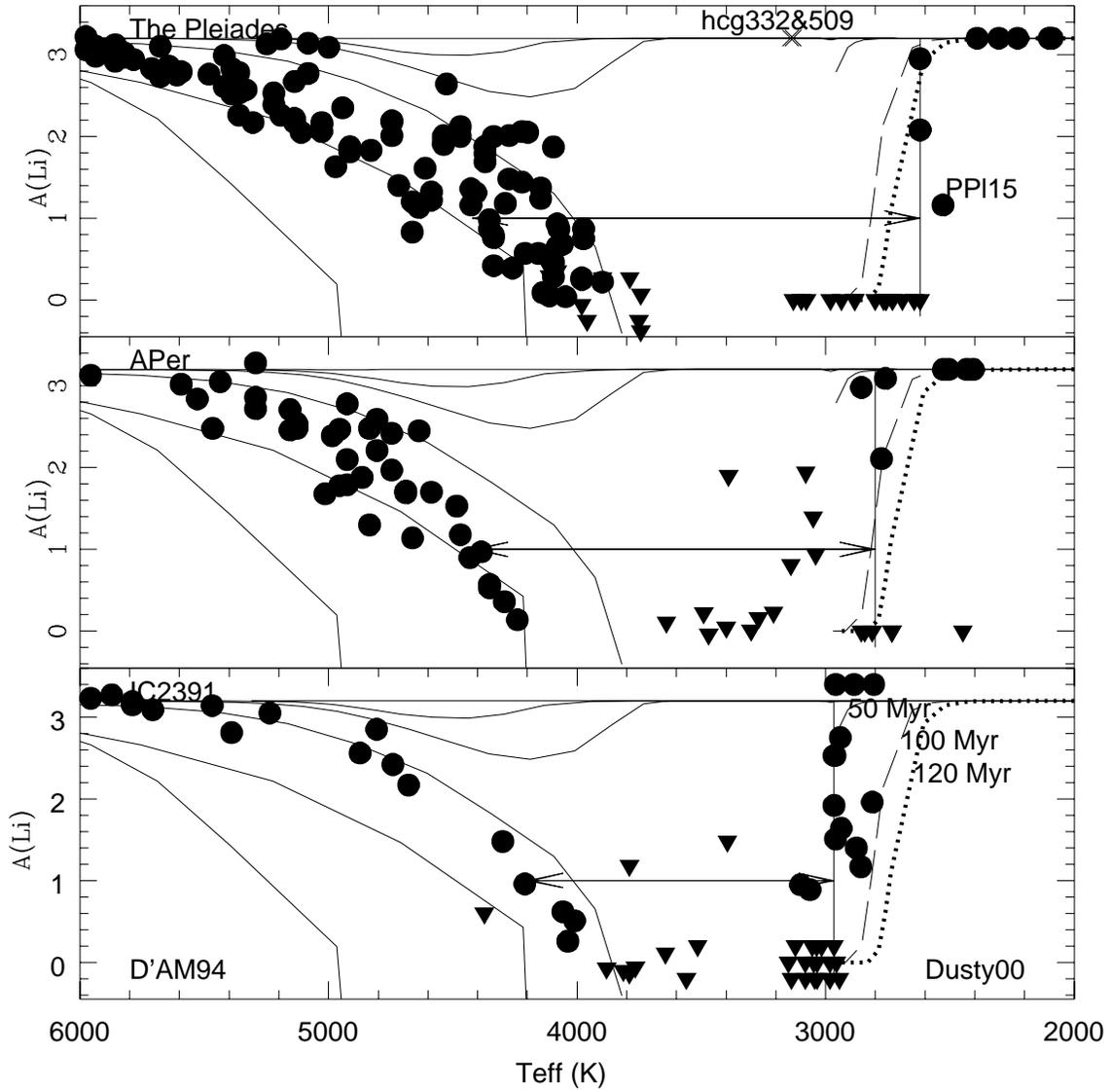}
 \caption{Lithium abundance versus effective temperature.
Actual  abundances and upper limits are shown as circles
and triangles, respectively. Several lithium depletion  isochrones
from D'Antona \& Mazzitelli (1994) --1, 3, 5, 10, 20 and 100 Myr, left-- and 
Chabrier et al. (2000) --50, 100 and 120 Myr; right-- are included.}
 \end{figure*}

\setcounter{figure}{11}
    \begin{figure*}
    \centering
    \includegraphics[width=16.2cm]{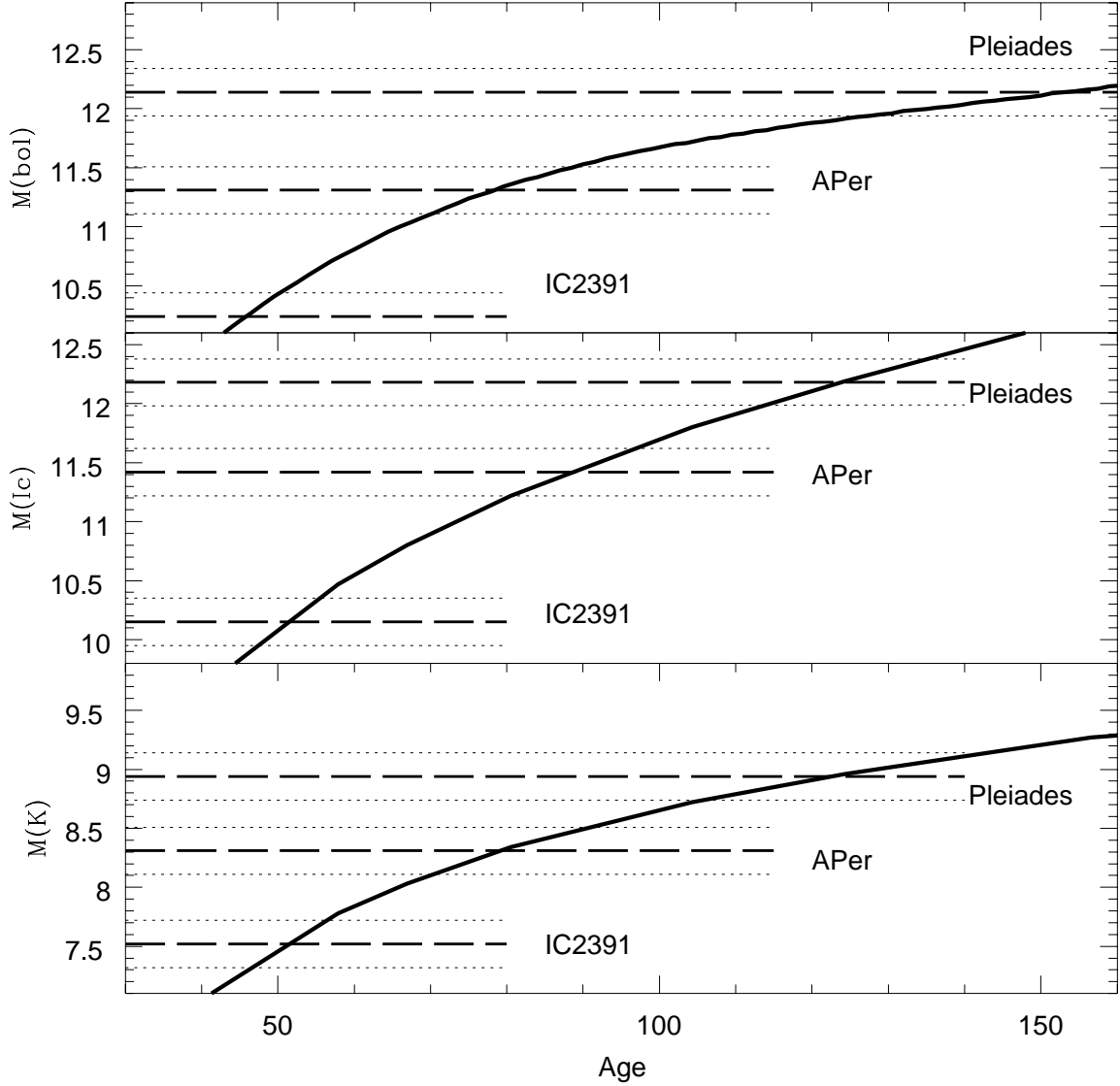}
 \caption{Location of the Lithium Depletion Boundary (LDB)
 for IC~2391, Alpha Per and the Pleiades.
For the upper panel we have used data from D'Antona \& Mazzitelli (1997),
 whereas the other two display results from I. Baraffe (priv. comm.).}
 \end{figure*}

\setcounter{figure}{12}
    \begin{figure*}
    \centering
    \includegraphics[width=16.2cm]{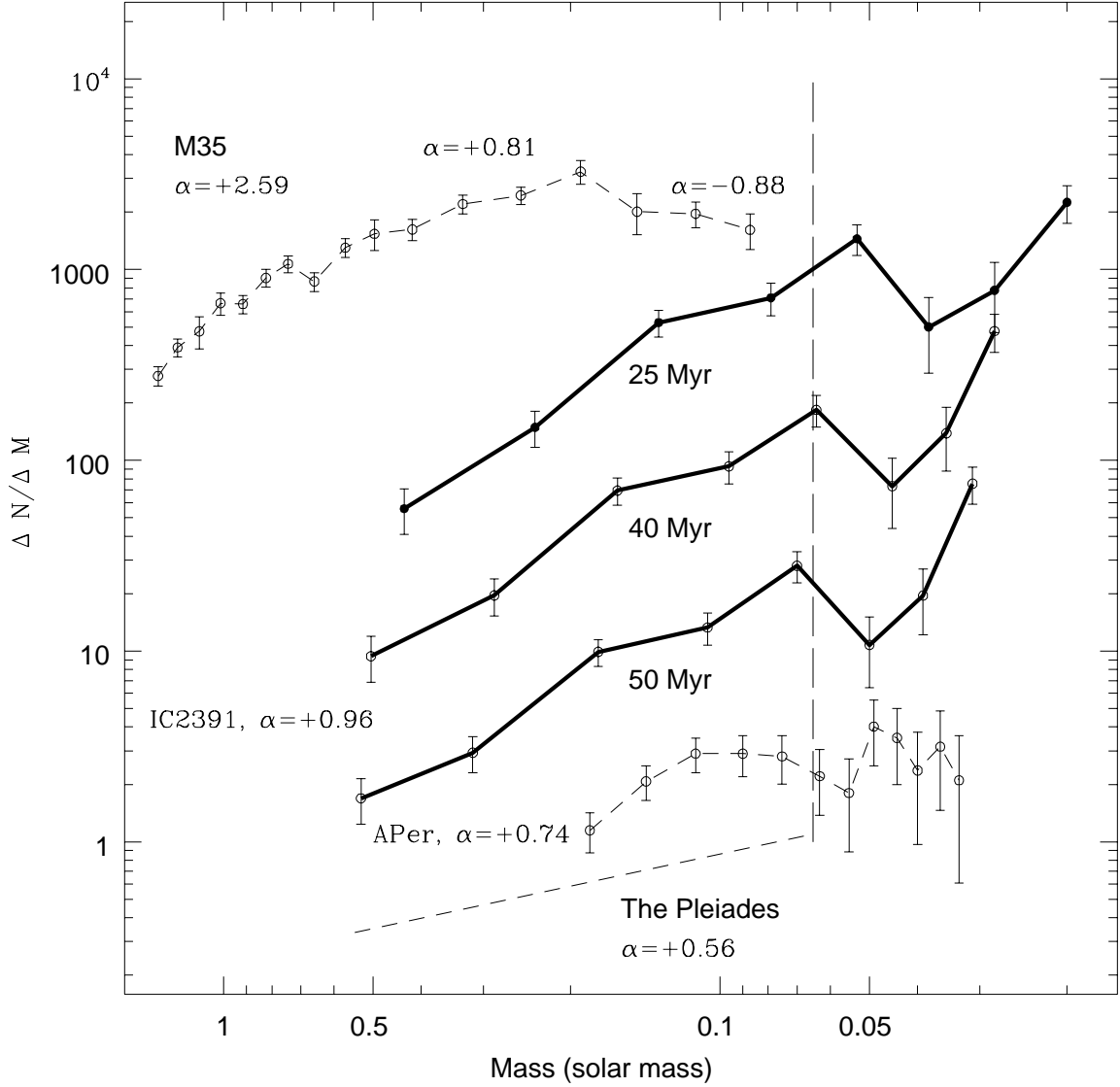}
 \caption{Mass Functions corresponding to several 
young open clusters. Those corresponding to IC~2391, computed assuming 
different ages, are highlighted with a thick, solid line. The vertical 
segment (long-dashed) represent the completeness limit for our 
survey. }
 \end{figure*}

\setcounter{table}{0}
\begin{table*}
\footnotesize
\caption[]{Photometry for our IC~2391 targets. }
\begin{tabular}{lc rrrrrrl}
\hline
 Name   &   RA (2000.0) DEC      &   V   &  $Ic$  & $R-I$  &     $J$  & $H$    &  $Ks$   &    Run$^1$                 \\ %
                                                                                                                             
\hline                                                                                                                       
VXRP39a  & 8 41 51.54 -53 20 59.7 & 15.74 & 13.30  & 1.34   &   12.049 & 11.392 & 11.179  &   CTIO/HYDRAII          \\ %
VXRP71a  & 8 44 19.08 -53 08 28.9 & 15.32 & 12.91  & 1.32   &   11.672 & 11.069 & 10.795  &   CTIO/R\&C             \\ %
PP02     & 8 39 05.67 -53 21 44.6 & 17.15 & 14.22  & 1.65   &   12.723 & 12.098 & 11.848  &   CTIO/HYDRAII          \\ %
PP07     & 8 39 29.58 -53 21 04.4 & 17.31 & 14.11  & 1.77   &   12.480 & 11.883 & 11.615  &   CTIO/HYDRAII          \\ %
PP11     & 8 44 04.65 -53 00 01.7 & 18.48 & 15.30  & 1.79   &   13.684 & 13.075 & 12.810  &   CTIO/R\&C             \\ %
PP14     & 8 40 30.31 -53 11 30.9 & 19.22 & 15.83  & 1.87   &   14.184 & 13.589 & 13.294  &   CTIO/R\&C,CTIO/HYDRAII\\ %
CTIO-002 & 8 35 44.88 -53 25 55.7 & --    & 17.157 & 2.260  &   16.367 & 14.234 & 14.035  &   Magellan/B\&C         \\ %
CTIO-012 & 8 36 45.72 -53 11 32.6 & 18.78 & 15.553 & 1.672  &   13.935 & 13.251 & 12.992  &   Magellan/B\&C         \\ %
CTIO-017 & 8 37 11.46 -52 36 35.7 & 19.30 & 15.989 & 1.752  &   14.461 & 13.860 & 13.587  &   CTIO/R\&C             \\ %
CTIO-026 & 8 37 18.19 -52 55 56.9 & --    & 16.352 & 2.001  &   14.507 & 13.880 & 13.492  &   Magellan/MIKE         \\ %
CTIO-038 & 8 37 59.20 -53 21 55.4 & 19.90 & 16.291 & 1.910  &   14.455 & 13.869 & 13.643  &   CTIO/R\&C,CTIO/HYDRAII\\ %
CTIO-041 & 8 38 11.89 -52 22 51.4 & 20.46 & 16.554 & 1.923  &   14.630 & 14.088 & 13.741  &   CTIO/R\&C             \\ %
CTIO-046 & 8 38 25.10 -53 19 10.9 & 18.92 & 15.958 & 1.697  &   14.420 & 13.803 & 13.435  &   CTIO/HYDRAII          \\ %
CTIO-049 & 8 38 27.15 -53 25 10.4 & 19.05 & 15.566 & 1.944  &   13.928 & 13.336 & 13.002  &   CTIO/HYDRAII          \\ %
CTIO-054 & 8 38 36.10 -53 25 52.1 & --    & 20.629 & 2.542  &    --    &  --    &  --     &   CTIO/HYDRAII          \\ %
CTIO-056 & 8 38 38.80 -53 07 57.5 & --    & 16.711 & 1.837  &   15.314 & 14.631 & 14.365  &   CTIO/HYDRAII          \\ %
CTIO-058 & 8 38 42.34 -53 29 31.3 & --    & 19.919 & 2.661  &    --    &  --    &  --     &   CTIO/HYDRAII          \\ %
CTIO-059 & 8 38 44.03 -53 22 51.0 & 19.47 & 16.050 & 1.819  &   14.460 & 13.757 & 13.526  &   CTIO/R\&C             \\ %
CTIO-061 & 8 38 47.07 -52 14 56.4 & --    & 17.309 & 2.141  &   15.274 & 14.677 & 14.206  &   CTIO/R\&C             \\ %
CTIO-062 & 8 38 47.30 -52 44 32.7 & 20.84 & 16.765 & 2.000  &   14.954 & 14.389 & 13.989  &   CTIO/R\&C             \\ %
CTIO-067 & 8 38 56.19 -52 51 38.1 & --    & 17.111 & 2.285  &   16.087 & 15.626 & 15.795  &   Magellan/B\&C         \\ %
CTIO-073 & 8 39 32.06 -53 28 12.7 & --    & 20.322 & 2.620  &    --    &  --    &  --     &   CTIO/HYDRAII          \\ %
CTIO-074 & 8 39 40.59 -53 06 07.7 & --    & 15.876 & 1.786  &   14.191 & 13.526 & 13.278  &   CTIO/HYDRAII          \\ %
CTIO-076 & 8 39 48.45 -53 13 58.5 & 18.47 & 15.278 & 1.681  &   13.664 & 13.045 & 12.745  &   CTIO/HYDRAII          \\ %
CTIO-077 & 8 40 09.53 -53 37 49.7 & 20.04 & 16.308 & 1.929  &   14.543 & 13.962 & 13.632  &   CTIO/R\&C,CTIO/HYDRAII\\ %
CTIO-081 & 8 40 14.77 -53 27 36.4 & 20.38 & 16.465 & 2.079  &   14.370 & 13.745 & 13.394  &   CTIO/R\&C,CTIO/HYDRAII\\ %
CTIO-083 & 8 40 16.07 -53 25 47.9 & 18.85 & 15.565 & 1.799  &   13.906 & 13.272 & 12.933  &   CTIO/HYDRAII          \\ %
CTIO-087 & 8 40 42.92 -53 09 19.0 & --    & 20.240 & 2.503  &    --    &  --    &  --     &   CTIO/HYDRAII          \\ %
CTIO-089 & 8 40 46.81 -53 13 52.1 & --    & 19.971 & 2.726  &    --    &  --    &  --     &   CTIO/HYDRAII          \\ %
CTIO-091 & 8 40 53.00 -52 23 00.4 & --    & 15.842 & 1.765  &   14.082 & 13.460 & 13.174  &   Magellan/B\&C         \\ %
CTIO-096 & 8 41 12.38 -53 09 10.3 & 19.75 & 16.144 & 1.912  &   14.344 & 13.818 & 13.404  &   CTIO/R\&C             \\ %
CTIO-097 & 8 41 26.00 -53 26 34.8 & --    & 15.087 & 1.751  &   13.456 & 12.761 & 12.541  &   CTIO/HYDRAII          \\ %
CTIO-098 & 8 41 29.18 -53 16 22.3 & 18.87 & 15.593 & 1.797  &   13.988 & 13.333 & 13.060  &   CTIO/HYDRAII          \\ %
CTIO-106 & 8 41 58.93 -53 12 36.4 & --    & 16.454 & 1.983  &   14.655 & 13.997 & 13.676  &   CTIO/R\&C             \\ %
CTIO-113 & 8 42 18.71 -52 39 40.1 & 21.9  & 17.282 & 2.135  &   15.083 & 14.377 & 14.030  &   CTIO/R\&C             \\ %
CTIO-136 & 8 43 15.15 -52 58 23.0 & 19.72 & 15.868 & 1.978  &   14.134 & 13.482 & 13.129  &   CTIO/R\&C             \\ %
CTIO-145 & 8 43 23.67 -53 14 16.9 & --    & 16.936 & 2.135  &   15.059 & 14.386 & 13.962  &   CTIO/R\&C             \\ %
CTIO-152 & 8 43 38.42 -52 50 55.6 & 18.11 & 14.891 & 1.781  &   13.337 & 12.714 & 12.452  &   CTIO/R\&C             \\ %
CTIO-160 & 8 44 02.10 -52 44 10.7 & 21.05 & 17.151 & 2.090  &   15.115 & 14.468 & 14.103  &   Magellan/B\&C         \\ %
CTIO-192 & 8 45 02.58 -52 59 28.8 & --    & 16.286 & 1.970  &   14.460 & 13.853 & 13.527  &   Magellan/MIKE         \\ %
CTIO-195 & 8 45 45.60 -53 12 37.8 & --    & 16.353 & 1.869  &   14.566 & 13.986 & 13.611  &   Magellan/MIKE         \\ %
CTIO-202 & 8 46 26.27 -53 01 53.5 & --    & 16.391 & 2.057  &   14.472 & 13.962 & 13.551  &   Magellan/MIKE         \\ %
CTIO-205 & 8 47 03.47 -52 46 52.3 & 20.21 & 16.211 & 2.065  &   14.217 & 13.744 & 13.350  &   CTIO/R\&C             \\ %
CTIO-206 & 8 40 40.84 -53 13 31.9 & --    & 15.658 & 1.730  &   13.697 & 13.132 & 12.785  &   CTIO/HYDRAII          \\ %
\hline
\end{tabular}
$\,$\\
$^1$ CTIO/R\&C.-     January 1999 (Paper I);
     CTIO/HYDRAII.-  March 10-13th, 1999;
     Magellan/B\&C.- March 11th, 2003;
     Magellan/MIKE.- December 11-14th, 2002.
\end{table*}

\setcounter{table}{1}
\begin{table*}
\small
\footnotesize
\caption[]{Spectroscopy data for IC~2391 candidate members.  }
\begin{tabular}{ll cccc cccc}
\hline
 Name    &  Ic     & W(Ha)$^1$       &W(Na)$^2$      &W(Li)$^2$  & A(Li)    & Teff$^7$ & Teff$^8$& Spectral&Member? \\
         &         & (\AA)           &    (\AA)       &    (\AA)       &          & (K)  & (K)   & type    &        \\          
\hline         
VXRP39a  & 13.30   &      3.8$\pm$0.3&  4.3$\pm$0.1  & $<$0.03         & $<$0.00 & 3515 & 3250 &  M3.5 &Yes\\  %
VXRP71a  & 12.91   &      4.9$\pm$0.3&  --           & $<$0.10         & $<$0.00 & 3560 & 3350 &  M3.0 &Yes\\  %
PP02     & 14.22   &      8.9$\pm$0.6&  3.5$\pm$0.6  & $<$0.04         & $<$0.00 & 3149 & 3075 &  M4.5 &Yes\\  %
PP07     & 14.11   &      7.1$\pm$0.7&  5.0$\pm$0.2  &    0.23$\pm$0.09&    0.89 & 3063 & 3000 &  M5.0 &Yes?\\  %
PP11     & 15.30   &      5.4$\pm$0.5&  4.7$\pm$0.2  & $<$0.06         & $<$0.00 & 3044 & 3075 &  M4.5 &Yes\\  %
PP14     & 15.83 &1.4/6.3$\pm$0.2/0.3&  5.4$\pm$0.2  & $<$0.03         & $<$0.00 & 2983 & 3000 &  M5.0 &Yes\\  %
CTIO-002 & 17.157  &     -3.0        &  out$^5$      &    --           &   --    & 2645 &  --  &  K    & NM\\  %
CTIO-012 & 15.553  &     13.4$\pm$0.7&  out$^5$      & $<$0.20         & $<$0.00 & 3122 & 3000 &  M5.0 &Yes\\  %
CTIO-017 & 15.989  &     11.0$\pm$1.1&  5.6$\pm$0.2  & $<$0.12         & $<$0.00 & 3081 & 3150 &  M4.0 &Yes\\  %
CTIO-026 & 16.352  &     19.9$\pm$2.5&  out$^5$      &    0.37$\pm$0.08&    1.17 & 2858 & 2900 &  M5.5 &Yes\\  %
CTIO-038 & 16.291&0.6/7.0$\pm$1.0/0.5&  5.0$\pm$0.5  &    0.53$\pm$0.09&    2.53 & 2966 & 3000 &  M5.0 &Yes\\  %
CTIO-041 & 16.554  &      8.7$\pm$0.8&  5.2$\pm$0.2  &    0.53$\pm$0.15&    2.53 & 2962 & 3000 &  M5.0 &Yes\\  %
CTIO-046 & 15.958  &   $<$0          &  7.3$\pm$0.2  & $<$0.07         & $<$0.00 & 3127 & 3075 &  M4.5 & NM\\  %
CTIO-049 & 15.566  &      0.4$\pm$0.2&  5.8$\pm$0.6  & $<$0.02         & $<$0.00 & 2956 & 3000 &  M5.0 &Yes\\  %
CTIO-054 & 20.629  &   $<$0          &  --           &    --           &   --    & --   &  --  &  --   & NM\\  %
CTIO-056 & 16.711  &     17.0$\pm$2.2&  5.6$\pm$0.6  &    --           &   --    & 3003 & 3000 &  M5.0 &Yes\\  %
CTIO-058 & 19.919  &   $<$0          &  --           &    --           &   --    & --   &  --  &  --   & NM\\  %
CTIO-059 & 16.050  & 34.1$^3$$\pm$1.6&  5.6$\pm$0.2  & $<$0.15         & $<$0.00 & 3017 & 3150 &  M4.0 &Yes\\  %
CTIO-061 & 17.309  &      2.3$\pm$1.0&  5.9$\pm$0.3  &    --           &   --    & 2801 & 2800 &  M6.0 &Yes\\  %
CTIO-062 & 16.765  &      7.2$\pm$1.0&  5.3$\pm$0.3  &    0.42$\pm$0.15&    1.64 & 2937 & 2800 &  M6.0 &Yes\\  %
CTIO-067 & 17.111  &     -2.5        &  out$^5$      &    --           &   --    & 2603 &  --  &  K    & NM\\  %
CTIO-073 & 20.322  &   $<$0          &  --           &    --           &   --    & --   &  --  &  --   & NM\\  %
CTIO-074 & 15.876  &      5.9$\pm$0.6&  6.4$\pm$1.0  & $<$0.04         & $<$0.00 & 3048 & 3000 &  M5.0 &Yes\\  %
CTIO-076 & 15.278  &     11.2$\pm$0.7&  4.9$\pm$0.3  & $<$0.02         & $<$0.00 & 3137 & 3000 &  M5.0 &Yes\\  %
CTIO-077 & 16.308&7.0/7.4$\pm$0.9/0.3&  5.5$\pm$0.3  & 1.2$\pm$0.4$^9$ &    3.4  & 2960 & 3000 &  M5.0 &Yes\\  %
CTIO-081 & 16.465&0.7/8.6$\pm$0.1/0.3&  6.0$\pm$0.4  &    0.82$\pm$0.15&    3.4  & 2885 & 2900 &  M5.5 &Yes$^6$\\  %
CTIO-083 & 15.565  &     10.9$\pm$0.3&  5.5$\pm$0.4  & $<$0.01         & $<$0.00 & 3035 & 3000 &  M5.0 &Yes\\  %
CTIO-087 & 20.240  &  em.$^4$        &  --           &    --           &   --    & --   &  --  &  --   &Yes?\\ %
CTIO-089 & 19.971  &  em.$^4$        &  --           &    --           &   --    & --   &  --  &  --   &Yes?\\ %
CTIO-091 & 15.842  &      9.9$\pm$0.8&  5.7$\pm$0.4  & $<$0.10         & $<$0.00 & 3035 & 3000 &  M5.0 &Yes\\  %
CTIO-096 & 16.144  &      8.7$\pm$0.7&  5.3$\pm$0.2  & $<$0.10         & $<$0.00 & 2966 & 3000 &  M5.0 &Yes\\  %
CTIO-097 & 15.087  &      7.3$\pm$1.3&  5.7$\pm$0.3  & $<$0.01         & $<$0.00 & 3082 & 3000 &  M5.0 &Yes\\  %
CTIO-098 & 15.593  &      2.0$\pm$0.2&  5.5$\pm$0.4  & $<$0.01         & $<$0.00 & 3037 & 3000 &  M5.0 &Yes\\  %
\hline  
\end{tabular}
\end{table*}

\setcounter{table}{1}
\begin{table*}
\small
\footnotesize
\caption[]{ (Continue)  }
\begin{tabular}{ll cccc cccc}
\hline
 Name    &  Ic     & W(Ha)$^1$       &W(Na)$^2$      &W(Li)$^2$  & A(Li)    & Teff$^7$ & Teff$^8$& Spectral&Member? \\
         &         & (\AA)           &    (\AA)       &    (\AA)       &          & (K)  & (K)   & type    &        \\          
\hline         
CTIO-106 & 16.454  &     14.1$\pm$1.8&  5.8$\pm$0.3  &    --           &   --    & 2944 & 2900 &  M5.5 &Possible\\  %
CTIO-113 & 17.282  &      4.6$\pm$1.0&  6.3$\pm$0.5  &    --           &   --    & 2812 & 2575 &  M7.0 &Yes\\  %
CTIO-136 & 15.868  &      6.1$\pm$0.9&  5.2$\pm$0.2  & $<$0.10         & $<$0.00 & 2945 & 2900 &  M5.5 &Yes\\  %
CTIO-145 & 16.936  &      7.9$\pm$2.9&  5.5$\pm$0.3  &    0.48$\pm$0.11&    1.96 & 2812 & 2800 &  M6.0 &Yes\\  %
CTIO-152 & 14.891  &      9.9$\pm$0.6&  5.1$\pm$0.3  & $<$0.02         & $<$0.00 & 3053 & 3150 &  M4.0 &Yes\\  %
CTIO-160 & 17.151  &     12.0$\pm$1.8&  out$^5$      &    0.9$\pm$0.2  &    3.4  & 2806 & 2575 &  M7.0 &Yes\\  %
CTIO-192 & 16.286  &      5.3$\pm$0.4&  out$^5$      &    0.40$\pm$0.15&    1.40 & 2876 & 3000 &  M5.0 &Yes\\  %
CTIO-195 & 16.353  &      9.8$\pm$1.5&  out$^5$      &    0.56$\pm$0.12&    2.75 & 2941 & 2900 &  M5.5 &Yes\\  %
CTIO-202 & 16.391  &      6.1$\pm$1.5&  out$^5$      &    --           &   --    & 2826 & 2900 &  M5.5 &Yes\\  %
CTIO-205 & 16.211  &      8.0$\pm$6.2&  5.2$\pm$0.6  &    --           &   --    & 2898 & 2900 &  M5.5 &Yes?\\  %
CTIO-206 & 15.658  &      7.4$\pm$0.8&  5.9$\pm$0.4  &    0.42$\pm$0.11&    0.95 & 3101 & 2900 &  M5.5 &Yes\\  
\hline  
\end{tabular}
$\,$\\
$^1$ W(H$\alpha$)$>$0 correspond to emission, whereas negative value correspond to absorption. \\
$^2$ All W(Na{\sc i}8200) and W(Li6709) are in absorption.\\
$^3$ Average of two observations in consecutive nights. The individual values are W(H$\alpha$)=45.5 and 18.8 \AA.\\
$^4$ In emission, with no continuum (very low S/N). \\
$^5$ Na{\sc i}(8200) out of range. \\
$^6$ Classified as ``possible'' member in Paper I. We have collected a higher S/N spectrum,\\ 
which indicates the presence of lithium. \\
$^7$ Teff from $(R-I)_C$ color. \\
$^8$ Teff from the spectral type. \\
$^9$ The detection of lithium is quite uncertain in this case. 
\end{table*}

\setcounter{table}{2}
\begin{table*}
\small
\caption[]{Summary of the lithium depletion boundary data (LDB).}
\begin{tabular}{l  ccc}
\hline
                    & IC~2391    & Alpha Per   & The Pleiades \\
\hline         
(m-M)o              & 5.95      & 6.23        & 5.60         \\
E(B-V)              & 0.06      & 0.096       & 0.04         \\
M($Ic$)             & 10.15     & 11.42       &  12.18       \\
M($Ms$)             &  7.52     &  8.31       &   8.94       \\
M(bol)              & 10.24     & 11.31       &  12.14       \\
Sp.Type             & M5        & M6.5        &  M6.5        \\
Teff LDB (K)        & 3050      & 2800        &  2650        \\
Turn-off age (Myr)  & 35        & 50          &  80--100     \\
LDB age      (Myr)  & 50$\pm$5  & 85$\pm$10   &  130$\pm$20  \\
Mass  (M$_\odot$)   &  0.12     &  0.085      &   0.075      \\
\hline  
\end{tabular}
$\,$\\
\end{table*}

\end{document}